\documentclass[12pt]{article}
\usepackage[pdftex]{graphicx}
\usepackage{url}
\usepackage{braket}
\usepackage{latexsym}
\usepackage{mathtools}
\usepackage{url}
\usepackage{graphicx}
\usepackage{bbm}
\usepackage{amsmath,amssymb,bm} 
\DeclareMathOperator{\Tr}{Tr}

\usepackage{amsmath}
\usepackage{cite}

\usepackage[utf8]{inputenc} 

\pdfoutput=1
\usepackage{tikz}
\usepackage{putex}
\usepackage{feyn}
\usepackage[vcentermath]{youngtab}
\usepackage{subfig}
\usepackage{lscape}

\usepackage{graphicx}
\usepackage{epstopdf}
\usepackage{enumerate}
\usepackage{cite}
\usepackage{tensor} 
\usepackage{slashed}
\usepackage{amsthm}
\usepackage{amsmath}
\usepackage{amssymb}
\usepackage{mathrsfs}
\usepackage{lgrind}
\usepackage{youngtab}

\usepackage{amsmath,amssymb,braket}

\usepackage{bbm}

\usepackage{hyperref}

\numberwithin{equation}{section}

\newcommand {\be} {\begin {equation}}
\newcommand {\ee} {\end {equation}}

\newcommand {\bes} {\begin {equation*}}
\newcommand {\ees} {\end {equation*}}

\def\CH{{\cal H}}

\newcommand{\beq}{\begin{equation}}
\newcommand{\eeq}{\end{equation}}

\def\be{ \begin{equation} }
\def\ee{ \end{equation} }

\def\Tr{{\operatorname{Tr}}}

\def \al{\alpha}

\def \be {\beta}

\def \beq { \begin{equation}}
\def \eeq {\end{equation}}
\def \pr {\partial}

\def \l {\left(}
\def \r {\right)}

\begin{document}


\institution{PU}{Department of Physics, Princeton University, Princeton, NJ 08544}
\institution{PCTS}{Princeton Center for Theoretical Science, Princeton University, Princeton, NJ 08544}
\institution{HU}{Department of Physics, Harvard University, Cambridge, MA 02138}

\title{
Spontaneous Breaking of $U(1)$ Symmetry in 
Coupled Complex SYK Models 
}

\authors{Igor R.~Klebanov,\worksat{\PU,\PCTS} Alexey Milekhin,\worksat{\PU}
Grigory Tarnopolsky,\worksat{\HU} Wenli Zhao\worksat{\PU}
}

\abstract{ As shown in \cite{Kim:2019upg}, two copies of the large $N$ Majorana SYK model
can produce spontaneous breaking of a $Z_2$ symmetry when they are coupled by appropriate quartic terms. In this paper we similarly study two copies of 
the complex SYK model coupled by a quartic term preserving the $U(1) \times U(1)$ symmetry. We also present a tensor counterpart of this coupled model.
When the coefficient $\alpha$ of the quartic term lies in a certain range, the coupled large $N$ theory is nearly conformal. We calculate the scaling dimensions of fermion bilinear operators as functions of $\alpha$. We
show that the operator $c_{1i}^\dagger c_{2i}$, which is charged under the axial $U(1)$ symmetry, acquires a complex dimension outside of the line of fixed points.  
We derive the large $N$ Dyson-Schwinger equations and show that, outside the fixed line, this $U(1)$ symmetry is spontaneously broken at low temperatures because this operator acquires an expectation value. We support these findings by exact diagonalizations extrapolated to large $N$. 
}

\date{}

\maketitle

\tableofcontents

\section{Introduction and summary}

There has been a great deal of interest in the fermionic quantum mechanical models which are exactly solvable in the large $N$ limit because they are dominated by a special class of Feynman diagrams, which are called melonic \cite{Bonzom:2011zz}. Perhaps the simplest such model is the Majorana SYK model consisting of a large number of Majorana fermions with random quartic interactions \cite{Kitaev:2015,Kitaev:2017awl}. Quantum mechanical models of this type have non-random tensor counterparts \cite{Witten:2016iux,Klebanov:2016xxf}, which have continuous symmetry groups
(for reviews of the melonic models see  \cite{Gurau:2011xp,Tanasa:2015uhr,Sarosi:2017ykf,Rosenhaus:2018dtp,Delporte:2018iyf,Klebanov:2018fzb,Gurau:2019qag,Trunin:2020vwy}). 
Both the random and non-random quantum mechanical models are
solvable via the same melonic Dyson-Schwinger (DS) equations \cite{Polchinski:2016xgd,Maldacena:2016hyu,Gross:2016kjj,Jevicki:2016bwu,Klebanov:2016xxf,Kitaev:2017awl}, which indicate that the model is nearly conformal. One can obtain richer dynamics when more than one Majorana SYK or tensor models are coupled \cite{Gu:2016oyy,Maldacena:2018lmt,Garcia-Garcia:2019poj,Kim:2019upg}. In particular, when two such models are coupled by certain quartic interactions with a coefficient $\alpha$, one finds a line of fixed points when $\alpha$ is positive, while a gapped $Z_2$ symmetry breaking phase appears when $\alpha$ is negative \cite{Kim:2019upg}. 

In this paper we make further progress in this direction by obtaining similar coupled models where a $U(1)$ symmetry is broken spontaneously in the large $N$ limit. 
Our starting point is the complex SYK model \cite{Sachdev:1992fk,Sachdev:2015efa,Davison:2016ngz,Gu:2019jub} (see also the earlier work \cite{Bohigas:1971vpj,French:1971et}), which has a $U(1)$ global symmetry. When two such models are coupled together by a quartic interaction preserving the $U(1)\times U(1)$ symmetry, 
\begin{align}
H =\sum_{i,j,k,l=1}^{N}J_{ij,kl} \Big (c^{\dag}_{1i}c^{\dag}_{1j}c_{1k}c_{1l} +
c^{\dag}_{2i }c^{\dag}_{2j}c_{2k}c_{2l} +8 \alpha c^{\dag}_{1i}c^{\dag}_{2j }c_{2k}c_{1l}\Big)\ ,
\label{ham1}
\end{align} 
we find that it is possible to break one of the $U(1)$ symmetries spontaneously. 
The phase where the $U(1)$ symmetry is broken by a VEV of operator $c_{1i}^\dagger c_{2i}$ is found for $\alpha<0$ and $\alpha>1$. In contrast with the breaking of discrete symmetry
in the coupled Majorana SYK model \cite{Kim:2019upg}, there is no gap in the full large $N$ spectrum due to the Nambu-Goldstone phenomenon. It manifests itself in splittings of order $1/N$ between the lowest states in different charge sectors. However, some specific charge sectors exhibit gaps of order $1$ above the ground state.  

We also exhibit a tensor counterpart of the coupled random model (\ref{ham1}) which consists of two coupled complex tensor models. The basic such model with $SU(N)^2\times O(N) \times 
U(1)$ symmetry was introduced in \cite{Klebanov:2016xxf}, and the two are coupled by an interaction which preserves the $SU(N)^2 \times O(N)\times
U(1)^2$ symmetry.\footnote{The meaning of $N$ in the tensor models is different from that in the SYK models.}

At the special coupling  $\alpha=1/4$, the $U(1)\times U(1)$ symmetry is enhanced to $U(2)\sim U(1)\times SU(2)$, and the Hamiltonian (\ref{ham1}) may be written compactly as 
 \begin{align}
H_{U(2)} =\sum_{i,j,k,l=1}^{N}J_{ij,kl} c^{\dag}_{\sigma i}c^{\dag}_{\sigma' j}c_{\sigma' k}c_{\sigma l} \,, \label{ham2}
\end{align} 
 where there is a sum over $\sigma,\sigma'= 1, 2$.
This is equal to the quartic term in the model of \cite{PhysRevB.72.045318}, which was argued to provide a description of quantum dots with irregular boundaries.
In (\ref{ham2}) the $U(1)$ is the usual charge symmetry, while the enhanced $SU(2)$ symmetry models the physical spin;
we may think of $\sigma$ as labeling the two spin states, up and down.

We note that some results on spontaneous $U(1)$ symmetry breaking in models with random couplings have already appeared in the literature \cite{Patel:2018rjp,
Wang:2019bpd,Esterlis:2019ola,2019arXiv190802757C,2019arXiv191207646C,Hauck_2020,abanov2020interplay,Wang:2020dtj}. For example, toy models of superconductivity
introduced in \cite{Wang:2019bpd,Esterlis:2019ola,Wang:2020dtj} include random Yukawa interactions of fermion-phonon type.

Other recently introduced models \cite{Patel:2018rjp,2019arXiv190802757C,2019arXiv191207646C} include random quartic couplings, as well as the non-random double-trace operator $O O^\dagger$, where $O$ is a "Cooper pair operator" $O\sim c_{i\uparrow} c_{i\downarrow}$. The models we study in this paper are somehwat different, and they appear to be the first examples of manifestly melonic theories where the spontaneous breaking of $U(1)$ symmetry can be established through analysis of the exact large $N$ Dyson-Schwinger equations.  

The structure of the paper is as follows. In section \ref{mm} we 
introduce some melonic models with $U(1)\times U(1)$ symmetry. 
They include a pair of coupled complex SYK models 
with Hamiltonian (\ref{ham1}), as well as the tensor 
counterpart of this model with Hamiltonian (\ref{tenmod}). 
In section \ref{cft} we discuss the symmetric saddle point of the large $N$ 
effective action, as well as fluctuations around it. 
There is a range $0\leq \alpha \leq 1$ where the symmetric saddle point is stable, 
while outside this fixed line a fermion bilinear 
operator, $c_{1i}^\dagger c_{2i}$, acquires a complex 
scaling dimension. In section \ref{DSeqs} we find a more general 
solution of the Dyson-Schwinger equations, which contains the off-diagonal 
Green's function $G_{12}$. It is stable outside the fixed line and 
indicates that the operator $c_{1i}^\dagger c_{2i}$ acquires an 
expectation value. This phase of the theory is characterized by the exponential fall-off of 
Green's functions at low temperatures. In section \ref{compressibility} we 
discuss the low-energy effective action in this phase and calculate the 
compressibility for the broken $U(1)$ degree of freedom. 
 In section \ref{EDsection} we support some of these results by 
Exact Diagonalizations at accessible values of $N$. Extrapolating the 
ground state energies and compressibilities to large $N$, 
we obtain good agreement with some of the results obtained using the DS equations. 
In section \ref{sec:su2} we present results for compressibilities at the special 
value $\alpha=1/4$ where the model has $U(2)$ symmetry. Some additional details can be found in the Appendices. 

As we were about to submit this paper to arXiv, we noticed the new paper \cite{2020arXiv200606019S} by S. Sahoo et al. where the model (\ref{ham1}) is also studied, with results similar to some of ours.

\section{Melonic models with $U(1)\times U(1)$ symmetry}
\label{mm}

In this section we 
introduce some melonic models with quartic Hamiltonians, which possess $U(1)\times U(1)$ symmetry. The first model with Hamiltonian (\ref{ham1}) consists of two copies of complex SYK model with a marginal $U(1)\times U(1)$ preserving interaction containing a dimensionless coupling, $\alpha$. 
We also formulate its tensor counterpart which has $SU(N)^2\times O(N)\times  
U(1)^2$ symmetry; it has the same Dyson-Schwinger equations as the random model.

\subsection{Two coupled complex SYK models}

Consider two sets of $ N$ complex fermions,  $c_{\sigma i}$,
where $\sigma=1,2$ and $i=1, \ldots ,N$:
\begin{equation}
\{ c^{\dag}_{\sigma i}, c_{\sigma' j} \}= \delta_{\sigma \sigma'} \delta_{ij} \ .
\end{equation}
The Hamiltonian coupling them is (\ref{ham1}), where 
$J_{ij,kl}$ is the random Gaussian complex tensor with zero mean $\overline{J_{ij,kl}}=0$; it satisfies $J_{ij,kl}=J^{*}_{kl,ij}$ in order for the Hamiltonian to be Hermitian.
We also assume anti-symmetry in the first and second pairs of indices: $J_{ij,kl}= - J_{ji,kl}=- J_{ij,lk}$.
 The variance is $\overline{|J_{ij,kl}|^{2}} = J^{2}/(2 N)^{3}$. 

So far the definition of the random tensor $J_{ij,kl}$ is incomplete. In fact, there is some freedom in its definition \cite{KKMT} even for the single complex SYK model. In this paper we will not use this freedom and will adopt the following minimal approach. 
We decompose $J_{ij,kl}$ as $J_{ij,kl}= \frac{1}{4}(T_{ij,kl}+T^{*}_{kl,ij})$, where $T_{ij,kl}$ is antisymmetric in the first and second pairs of indices and has no other symmetries. We then treat $T_{ij,kl}$ as  $N^{2}(N-1)^{2}/4$ independent complex Gaussian random variables, so $\overline{T_{ij,kl}}=0$ and  
$\overline{|T_{ij,kl}|^{2}} = J^{2}/N^{3}$.

The Hamiltonian (\ref{ham1}) has two $U(1)$ symmetries,
   \begin{align}
U(1)_{+}:& \quad c_{1i} \to e^{i\phi_+}c_{1i}, \quad c_{2i} \to e^{i\phi_+} c_{2i}\ ;\notag \\
U(1)_{-}:& \quad c_{1i} \to e^{i\phi_-}c_{1i}, \quad c_{2i} \to e^{-i\phi_-} c_{2i}\ .
\end{align} 
The corresponding conserved charges are $Q_\pm = Q_1 \pm Q_2$ where
\begin{equation} 
Q_1 = \frac 1 2 \sum_{i=1}^{N}  [c^{\dag}_{1i} ,  c_{1i}]\ , \qquad    Q_2 =\frac 1 2 \sum_{i=1}^{N}  [c^{\dag}_{2 i} ,  c_{2i}] 
\ .\end{equation}
The allowed values of $Q_1$ and $Q_2$ are $-\frac{N}{2}, -\frac{N}{2}+1, \ldots, \frac{N}{2}-1, \frac{N}{2}$; they are integer for even $N$ and half-integer for odd $N$.
Both $Q_+$ and $Q_-$ take integer values ranging from $- N$ to $N$. 
Also, there are constraints that $Q_+ + Q_-=2 Q_1$ and  $Q_+ - Q_-=2 Q_2$ are
even for $N$ even and odd for
$N$ odd.\footnote{We may consider a variant of the model where the $U(1)_+$ symmetry is gauged; in this case we have to restrict the Hilbert space to the sector with $Q_+=0$.}

The Hamiltonian also has the $Z_4$ symmetry 
\begin{equation}
c_{1i}\rightarrow c_{2i}\ , \qquad c_{2i}\rightarrow -c_{1i}\ , 
\label{zfour}
\end{equation}
which is analogous to the $Z_4$ symmetry which played an important role in \cite{Kim:2019upg}. Another important symmetry is the particle-hole symmetry
\begin{equation}
c_{1i}\leftrightarrow c_{1i}^\dagger \ , \qquad  c_{2i}\leftrightarrow c_{2i}^\dagger \ , \qquad  J_{ij,kl}\rightarrow J^*_{ij,kl}\ .
\label{phole}
\end{equation}
In order to make the Hamiltonian invariant under this symmetry for general $\alpha$, we have to add to it certain quadratic and c-number terms which are exhibited in (\ref{pholesymmetric}).\footnote{Note that, together with the symmetry which exchanges $c_1^i$ and $c_2^i$, the unitary discrete symmetries of (\ref{ham1}) are $D_4\times \mathbb{Z}_2.$ }

Note that, since the random coupling $J_{ij,kl}$ is complex, the $U(1)_+$ and $U(1)_-$ are on a different footing: the charge conjugation acting on the second flavor $c_{2i}$, 
\begin{equation}
C_{2}^{\dagger} c_{2i}C_{2}=c_{2i}^{\dagger} 
\label{chargeconj}
\end{equation}
is not a symmetry of the Hamiltonian (\ref{ham1}).  The $U(1)_+$ is the overall charge symmetry, while the ``axial" symmetry $U(1)_-$ may be thought of as a spatial rotation around the third axis.  
We will show that, for $\alpha<0$, the $U(1)_-$ may be broken spontaneously in the large $N$ limit, but the charge symmetry $U(1)_+$ remains unbroken. 
Holographically, the $U(1)_-$ has a simple physical meaning: a holographic state charged under $U(1)_-$ corresponds to bulk solutions with an electric field turned on.

Using the standard procedure for integrating over disorder and introducing bilocal fields,
\begin{equation}
G_{\sigma \sigma'}(\tau_{1},\tau_{2}) =\frac {1}  {N}\langle T c_{\sigma i} (\tau_{1}) c_{\sigma' i}^\dagger  (\tau_{2}) \rangle\ ,
\end{equation}
 and 
 $\Sigma_{\sigma \sigma'}(\tau_{1},\tau_{2})$, we write down the effective action 
    \begin{align}
&I = -\log\det (\delta'(\tau_{12})\delta_{\sigma\sigma'}-\Sigma_{\sigma \sigma'}(\tau_{1},\tau_{2})) -\int d\tau_{1}d\tau_{2}\Sigma_{\sigma \sigma'}(\tau_{1},\tau_{2})G_{\sigma' \sigma}(\tau_{2},\tau_{1}) -\frac{J^{2}}{4}\int d\tau_{1}d\tau_{2}V(G_{\sigma \sigma'})\,, \notag\\
&V(G_{\sigma \sigma'}) = G_{11}^{2}(\tau_{1},\tau_{2})G_{11}^{2}(\tau_{2},\tau_{1}) +G_{22}^{2}(\tau_{1},\tau_{2})G_{22}^{2}(\tau_{2},\tau_{1})+
2G_{12}^{2}(\tau_{1},\tau_{2})G_{21}^{2}(\tau_{2},\tau_{1})\notag\\
&\qquad +16\alpha\big(G_{11}(\tau_{1},\tau_{2})G_{11}(\tau_{2},\tau_{1})+G_{22}(\tau_{1},\tau_{2})G_{22}(\tau_{2},\tau_{1})\big)G_{12}(\tau_{1},\tau_{2})G_{21}(\tau_{2},\tau_{1}) +\notag\\
&\qquad +16\alpha^{2}\big(G_{11}(\tau_{1},\tau_{2})G_{22}(\tau_{1},\tau_{2})+G_{12}(\tau_{1},\tau_{2})G_{21}(\tau_{1},\tau_{2})\big)\big
(G_{11}(\tau_{2},\tau_{1})G_{22}(\tau_{2},\tau_{1})+G_{12}(\tau_{2},\tau_{1})G_{21}(\tau_{2},\tau_{1})\big)\,.
\label{effaction}
 \end{align}

 For $\alpha=1/4$ this can be nicely written as 
     \begin{align}
V(G_{\sigma \sigma'}) = \frac{1}{2} \textrm{Tr}\big(G(\tau_{1},\tau_{2})G(\tau_{2},\tau_{1})\big)^{2}+
\frac{1}{2} \textrm{Tr}\big(G(\tau_{1},\tau_{2})G(\tau_{2},\tau_{1})G(\tau_{1},\tau_{2})G(\tau_{2},\tau_{1})\big)\,,
   \end{align} 
   where  $G(\tau_{1},\tau_{2})$ is a $2\times 2$ matrix with elements $G_{\sigma\sigma'}(\tau_{1},\tau_{2})$. 
 So we can clearly see that $V(G_{\sigma \sigma'})$ is invariant under the global $U(2)$ transformations $G(\tau_{1},\tau_{2})\to U^{\dag} G(\tau_{1},\tau_{2})U$.

\subsection{Tensor counterpart of the random model}

Let us recall that the tensor counterpart of the standard complex SYK model \cite{Sachdev:2015efa,Gu:2019jub} is given by the tensor model with Hamiltonian \cite{Klebanov:2016xxf,Klebanov:2018fzb}
\begin{equation}
h= g  \bar \psi^{a_1 b_1 c_1} \bar \psi^{a_2 b_1 c_2} \psi^{a_1 b_2 c_2} \psi^{a_2 b_2 c_1}
\ .
\label{basictensor}
\end{equation}
The tensor indices range from $1$ to $N$, so that the model contains $N^3$ fermions, and the dimension of its Hilbert space is $2^{N^3}$.
The model has $SU(N)^2\times O(N) \times U(1)$ symmetry: the $O(N)$ symmetry acts on the second index of the tensor, while the two $SU(N)$ symmetries act on the first and third indices, respectively.  Exchanging the two $SU(N)$ groups 
changes $h\to -h$. 
  
Now we need to similarly determine the tensor counterpart of two coupled cSYK models (\ref{ham1}).
As we show in Appendix \ref{DiagSD}, the same Dyson-Schwinger equations as for this random model follow from the coupled tensor model with $SU(N)^2\times O(N)\times U(1)^2$ symmetry, which has the Hamiltonian
\begin{align}
H_{\rm tensor}= \frac{g}{2} \bigg (& \bar \psi_1^{a_1 b_1 c_1} \bar \psi_1^{a_2 b_1 c_2} \psi_1^{a_1 b_2 c_2} \psi_1^{a_2 b_2 c_1} +
\bar \psi_2^{a_1 b_1 c_1} \bar \psi_2^{a_2 b_1 c_2} \psi_2^{a_1 b_2 c_2} \psi_2^{a_2 b_2 c_1}\notag \\
+ & 4\alpha \left (\bar \psi_1^{a_1 b_1 c_1} \bar \psi_2^{a_2 b_1 c_2} \psi_2^{a_1 b_2 c_2} \psi_1^{a_2 b_2 c_1} - \bar \psi_1^{a_1 b_1 c_1} \bar \psi_2^{a_2 b_1 c_2} \psi_2^{a_2 b_2 c_1} \psi_1^{a_1 b_2 c_2} \right ) \bigg )\ .
\label{tenmod}
\end{align}
Under interchange of the two $SU(N)$ groups the Hamiltonian changes sign, and we have chosen the coupling term multiplied by $\alpha$ to preserve this discrete symmetry.
The $U(1)\times U(1)$ symmetry acts analogously to that in the random model, 
   \begin{align}
U(1)_{+}:& \quad \psi_1^{abc} \to e^{i\phi_+} \psi_1^{abc}, \qquad  \psi_2^{abc}  \to e^{i\phi_+} \psi_2^{abc} \ ;\notag \\
U(1)_{-}:&\quad \psi_1^{abc} \to e^{i\phi_-} \psi_1^{abc}, \qquad  \psi_2^{abc}  \to e^{-i\phi_-} \psi_2^{abc}  \ .
\end{align}
The Hamiltonian is also symmetric under the $\pi/2$ rotation $\psi_1^{abc}\rightarrow \psi_2^{abc}$, $\psi_2^{abc}\rightarrow -\psi_1^{abc}$.

In the tensor model (\ref{tenmod}) we may gauge the non-abelian symmetry $SU(N)^2\times O(N)$, restricting the states and operators to the sector invariant under this symmetry. 
Furthermore, as in the random counterpart (\ref{ham1}), it is possible to gauge the $U(1)_+$ symmetry.  

For $\alpha=1/4$, the symmetry is enhanced to $SU(N)^2\times O(N)\times U(2)$, and the Hamiltonian may be written as
\begin{align}
H_{\rm tensor}=\frac{g}{4}\left (\bar\psi_\sigma^{a_1b_1c_1}\bar\psi_{\sigma'}^{a_2b_1c_2}\psi_{\sigma'}^{a_1b_2c_2}\psi_\sigma^{a_2b_2c_1}
- \bar\psi_\sigma^{a_1b_1c_1}\bar\psi_{\sigma'}^{a_2b_1c_2}\psi_{\sigma'}^{a_2b_2c_1} \psi_\sigma^{a_1b_2c_2} \right )\ .
\end{align} 

For $\alpha=0$, the Hamiltonian (\ref{tenmod}) becomes a sum of two Hamiltonians (\ref{basictensor}).
In the tensor model (\ref{tenmod}) the gauged $SU(N)$ symmetries forbid correlators of the form $\langle \psi_\sigma^{abc}(t) \psi_{\sigma'}^{a'b'c'}(0)\rangle $, and the corresponding operators 
$\psi_\sigma^{abc} \partial_t^{m} \psi_{\sigma'}^{abc}$ 
are not allowed 
(in the random model, these operators do not receive ladder corrections). The symmetries do allow correlators of the form $\bar \psi_\sigma^{abc} \partial_t^{m} 
\psi_{\sigma'}^{abc}$, and their
large $N$ scaling dimensions are non-trivial. We will determine their values as functions of $\alpha$ in the next section, and show that one of them is complex for $\alpha<0$ and 
$\alpha>1$. 

 \section{Scaling Dimensions of Fermion Bilinears}
\label{cft}

 First let us study the large $N$ saddle point where
     \begin{align}
G_{12}=G_{ 21}=0\, \quad  \Sigma_{12}=\Sigma_{ 21}=0\, ,
   \end{align} 
  so that the $U(1)_{+}\times U(1)_{-}$ symmetry is preserved.   
Next it is reasonable to assume that $G_{11}(\tau_{1},\tau_{2})=G_{22}(\tau _1,\tau _2)=G(\tau_{12})$, where $G(\tau)$ is the particle-hole symmetric Green's function, so $G(-\tau)=-G(\tau)$. And we obtain 
   \begin{align}
\label{sym:J}
&\partial_{\tau}G(\tau) -\int d\tau' \Sigma(\tau -\tau')G(\tau')= \delta(\tau)\,, \notag\\
&\Sigma(\tau) = J^{2}(1+8\alpha^{2})G^{3}(\tau)\,,
 \end{align} 
 which is the standard SYK Dyson-Schwinger (DS) equations with $J' =J\sqrt{1+8\alpha^{2}} $. We will see that this diagonal saddle point describes the theory in the range 
$0\leq \alpha \leq 1$, where various large $N$ quantities are related to those in the $\alpha=0$ theory by the rescaling of $J$. For example, the ground state energy is
\begin{equation}
E_0(\alpha) = 2 E_0^{\textrm{cSYK}} \sqrt{ 1+8\alpha^{2} }\approx - 0.1624 N J \sqrt{ 1+8\alpha^{2} }\ .
\end{equation}

Now we consider the bilinear spectrum at the nearly conformal saddle \ref{sym:J}. They can be obtained by considering the melonic Bethe-Salpeter equations for the three point functions. Due to $U(1)_+\times U(1)_-$ symmetry, we can separate the computations in terms of 
\begin{equation}
v_{\sigma \sigma'}^{0,2(\sigma-\sigma')}(\tau,0, \infty)=\langle c^{\dagger}_{\sigma i}(\tau)c_{\sigma' i}(0) \mathcal{O}_h^{0, 2(\sigma-\sigma')}(\infty) \rangle\ , 
\end{equation}
between elementary fermions $c_{\sigma i}^{\dagger}, c_{\sigma' i}$ and a primary operator $\mathcal{O}_h^{0,2(\sigma-\sigma')}$ with dimension $h$ and $U(1)_+\times U(1)_-$ charge $(0,2(\sigma -\sigma')).$ Note that operators with non-zero $U(1)_+$ charge do not receive ladder correction in the large $N$ limit due to $J_{ij,kl}$ being complex.
 
It is convenient to write down schematically the bilinear operators $\{\mathcal{O}_{m,+}^{0,0}, \mathcal{O}^{0,0}_{m,-}, \mathcal{O}_{m}^{0,2}, \mathcal{O}_{m}^{0,-2}\}$: 
 \begin{equation}
 \mathcal{O}^{0,0}_{m,\pm}=c_{1i}^{\dagger}\partial_\tau^m c_{1i} \pm c_{2i}^{\dagger}\partial_\tau^m c_{2i},  
\quad \mathcal{O}^{0,2}_{m}= c_{1i}^{\dagger}\partial_\tau^m c_{2i}+(-1)^{m+1} c_{2i}\partial_\tau^m c_{1i}^{\dagger},  
\quad\mathcal{O}_m^{0,-2}= c_{2i}^{\dagger}\partial_\tau^m c_{1i}+(-1)^{m+1} c_{1i}\partial_\tau^m c_{2i}^{\dagger}\ .
 \end{equation}
This simple form of operators applies to the free UV theory (for a more precise form of the conformal primary operators in the SYK model, see \cite{Gross:2017hcz}), but
in the interacting IR theory the operators have a more complicated form. We will now present a calculation of their scaling dimensions which is exact in the IR limit of the large $N$ theory.

 For $(0,0)$ operators the scaling dimensions are determined by the following matrix: 
\begin{equation}\label{u1n}
K_{(0,0)}=\frac{1}{1+8\alpha^2}\begin{pmatrix}\frac{2}{3}K_c-\frac{1}{3}K_c^T+\frac{8\alpha^2}{3}K_c& \frac{8\alpha^2}{3}(K_c-K_c^T)\\ \frac{8\alpha^2}{3}(K_c-K_c^T) & \frac{2}{3}K_c-\frac{1}{3}K_c^T+\frac{8\alpha^2}{3}K_c\end{pmatrix}, 
\end{equation}
where we define $K_c$ as the conformal kernel of a single SYK/tensor model with Majorana fermions: 
\begin{equation}
K_c(\tau_{1},\tau_{2},\tau_{3},\tau_{4})=-\frac{3}{4\pi}\frac{\text{sgn}(\tau_{13})\text{sgn}(\tau_{24})}{|\tau_{13}|^{2\Delta}|\tau_{24}|^{2\Delta}|\tau_{34}|^{2-4\Delta}}\ , \qquad \Delta=\frac{1}{4}\ .
\end{equation}
which has eigenvalues in the anti-symmetric and symmetric sectors as $g_a(h),3g_s(h),$ with
\begin{equation}
g_a(h)=-\frac{3}{2}\frac{\tan(\frac{\pi}{2}(h-\frac{1}{2}))}{h-\frac{1}{2}}\ , \qquad 
g_s(h)=-\frac{1}{2}\frac{\tan(\frac{\pi}{2}(h+\frac{1}{2}))}{h-\frac{1}{2}}\ ,
\end{equation}
 and $K_c^T(\tau_1,\tau_2,\tau_3,\tau_4)=K_c(\tau_1,\tau_2,\tau_4,\tau_3).$ 
 
 For the $(0,\pm 2)$ operators, they have the same anomalous dimensions determined by 
\begin{equation}
K_{(0,\pm2)}=\frac{ 8\alpha K_c-8\alpha^2K_c^T}{3(1+8\alpha^2)}\,.
\end{equation}
 
 As a result, the scaling dimensions of the bilinear operators $\{\mathcal{O}_{m,+}^{0,0}, \mathcal{O}^{0,0}_{m,-}, \mathcal{O}_{m}^{0,2}, \mathcal{O}_{m}^{0,-2}\}$ are determined by equating to $1$ the following functions: 
  \begin{align}\label{spec}
  &\{g_{a}(h),\frac{3-8\alpha^{2}}{3(1+8\alpha^{2})}g_{a}(h), \frac{8\alpha(\alpha+1)}{3(1+8\alpha^{2})}g_{a}(h), \frac{8\alpha(\alpha+1)}{3(1+8\alpha^{2})}g_{a}(h)\} \quad  m \text{ odd}\ ,\notag \\&
  \{g_{s}(h), g_s(h), \frac{8\alpha(1-\alpha)}{1+8\alpha^{2}}g_{s}(h), \frac{8\alpha(1-\alpha)}{1+8\alpha^{2}}g_{s}(h)\}\quad  m \text{ even} \ .
  \end{align}
 The series of scaling dimensions coming from solving $g_a(h)=1$ and $g_s(h)=1$ are the same as those found in a single complex SYK model or 
the $SU(N)^2\times O(N)\times U(1)$ tensor model \cite{Klebanov:2016xxf}.  
Thus, for any $\alpha\neq \frac 1 4$, there are two $h=1$ modes corresponding to the $U(1)\times U(1)$ symmetry.

 For $\alpha=\frac 1 4$, we find 
  \begin{align}
  \{g_{a}(h), \frac{5}{9}g_{a}(h),  \frac{5}{9}g_{a}(h), \frac{5}{9}g_a(h)\} , \quad 
  \{g_{s}(h), g_{s}(h), g_s(h), g_s(h)\}\, .
  \end{align}
Thus, four modes with $h=1$ are present. They are solutions with the smallest dimensions in their series and correspond to operators 
\begin{equation}
c_{1i}^{\dagger}c_{1i} \pm c_{2i}^{\dagger}c_{2i} ,  \qquad c_{1i}^{\dagger}c_{2i}\pm c_{2i}^{\dagger}c_{1i}    
\ ,
\end{equation}
which are proportional to the generators of the $U(2)$ symmetry $\frac{1}{2}c_{\sigma i}^{\dag}\lambda^{a}_{\sigma \sigma'}c_{\sigma' i}$. 

 ln contrast to the coupled Majorana SYK model \cite{Kim:2019upg}, the large $N$ operator spectrum (\ref{spec}) does not exhibit a duality symmetry. A duality (\ref{fakedual} can be explored at level of DS equations after assuming certain symmetries on the correlators, but fluctuations not obeying such symmetries prevent this duality from being exact. 
For example, the theory at $\alpha=1$ is not equivalent to that at $\alpha=0$. For $\alpha=1$, we note that
 the operator 
$\mathcal{O}^{0,0}_{1,-} = c_{1i}^{\dagger} \partial_\tau c_{1i} - c_{2i}^{\dagger} \partial_\tau c_{2i}$
has dimension $h \approx 1.2829$. Since this lies in the range 
$1< h <\frac{3}{2}$, the conformal solution might not be described by a Schwarzian theory \cite{Maldacena:2016upp}. In fact, $\mathcal{O}_{1,-}^{0,0}$ has scaling dimension in this range when $\alpha>\sqrt{\frac{3}{8}}$.

\begin{figure}[h!]
  \begin{center}  
    \includegraphics [width=0.5\textwidth, angle=0.]{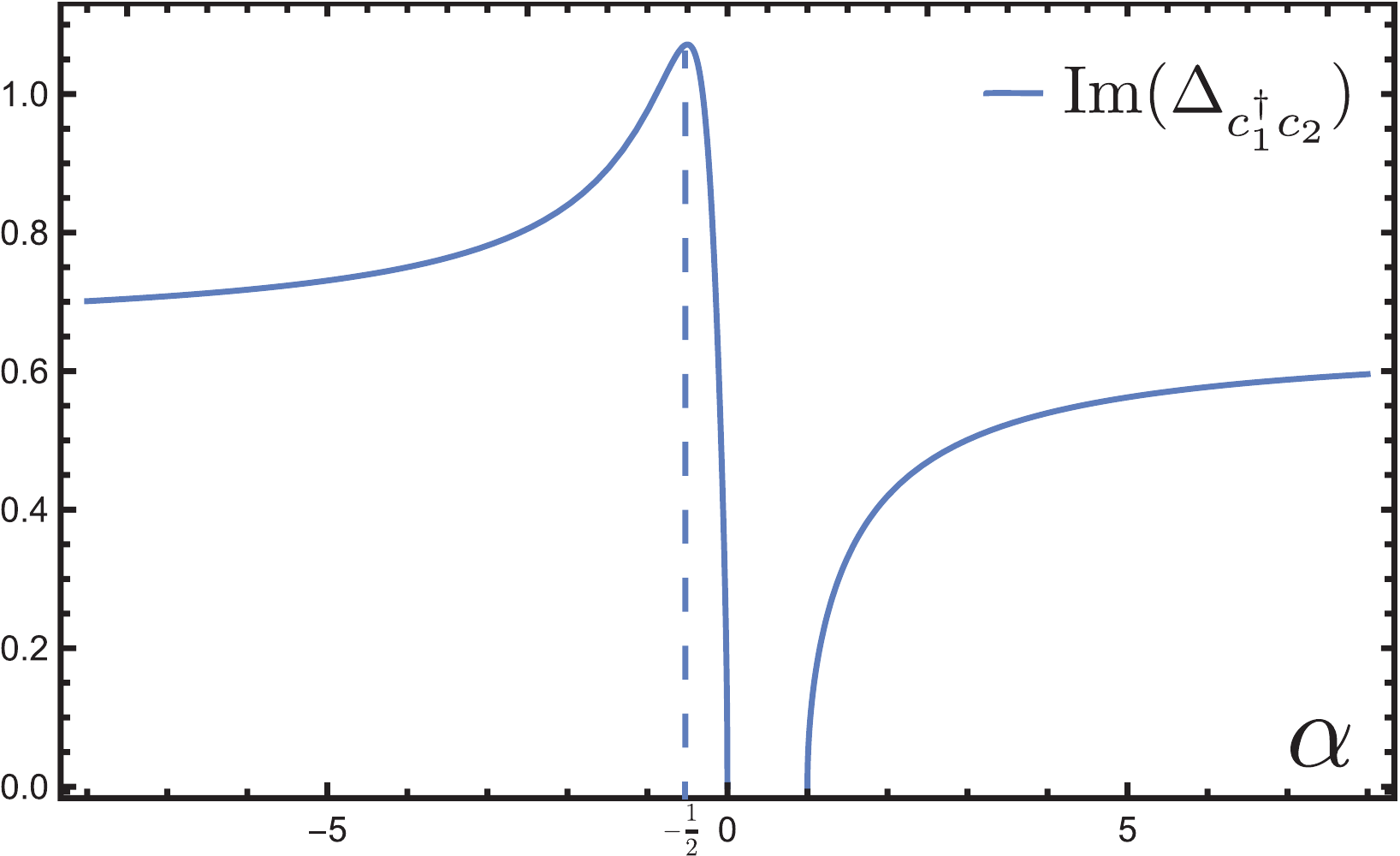}
  \end{center}
  \caption{The imaginary part of the scaling dimension of operator $c^\dagger_{1i} c_{2i}$. It reaches its maximum at $\alpha=-1/2$.} 
\label{scaldimalpha} 
\end{figure}

 For $\alpha<0$ or $\alpha >1$ the nearly conformal phase becomes unstable because the scaling dimension of operators ${O}^{0,\pm 2}_{0}$ becomes complex. 
The plot of its imaginary part as a function of $\alpha$ is in fig. \ref{scaldimalpha}. We note that it reaches its maximum when $\alpha=-1/2.$ The antisymmetric sector cannot 
 have such an instability for any $\alpha$ since $ -1/3 <\frac{3-8\alpha^{2}}{3 (1+8\alpha^{2})} \leq 1$ and$ -1/3 < \frac{ 8\alpha  (\alpha +1)}{3 \left(1+8\alpha ^2\right)}\leq 2/3$. So the lower bound is greater than $1/k_{a}(1/2)=-4/(3\pi)$.  In such cases, the real infrared solution acquires VEV of ${O}^{0,\pm 2}_{0}$  corresponding to the spontaneous breaking of $U(1)_-$ symmetry.

\section{General Dyson-Schwinger equations and their numerical solution}\label{DSeqs}

In this section we study the DS equations more generally and show that, for $\alpha<0$ or $\alpha >1$, the solution with lowest free energy breaks the $U(1)_-$ symmetry.
These equations may be obtained by varying the effective action (\ref{effaction}). The first series is
     \begin{align}\label{firstsdeq}
\partial_{\tau_{1}}G_{\sigma \sigma'}(\tau_{1},\tau_{2}) -\int d\tau_{3} \Sigma_{\sigma \sigma''}(\tau_{1},\tau_{3})G_{\sigma'' \sigma'}(\tau_{3},\tau_{2})= \delta_{\sigma \sigma'}\delta(\tau_{12})\,.
   \end{align} 
 For the second series we find 
      \begin{align}\label{secondsdeq}
\Sigma_{11}(\tau_{12}) =&-J^2 G_{11}(\tau_{12})^2 G_{11}(\tau_{21})-4\alpha  J^2 G_{11}(\tau_{12}) \big(G_{12}(\tau_{12}) G_{21}(\tau_{21})+G_{12}(\tau_{21}) G_{21}(\tau_{12})\big) \notag\\
&-8\alpha ^2 J^2 G_{22}(\tau_{21}) \big(G_{11}(\tau_{12}) G_{22}(\tau_{12})+G_{12}(\tau_{12}) G_{21}(\tau_{12})\big)\,, \notag\\
\Sigma_{12}(\tau_{12}) =&-J^2 G_{12}(\tau_{12})^2 G_{21}(\tau_{21})
-4 \alpha  J^2 G_{12}(\tau_{12}) \big(G_{11}(\tau_{12}) G_{11}(\tau_{21})+G_{22}(\tau_{12}) G_{22}(\tau_{21})\big) \notag\\
&-8 \alpha ^2 J^2 G_{12}(\tau_{21}) \big(G_{11}(\tau_{12}) G_{22}(\tau_{12})+G_{12}(\tau_{12}) G_{21}(\tau_{12})\big)  \ .
\end{align} 
The equation for $\Sigma_{22}(\tau_{12})$ is obtained from $\Sigma_{11}(\tau_{12})$ by $G_{11}\leftrightarrow G_{22}$, and that for 
$\Sigma_{21}(\tau_{12})$ is obtained from $\Sigma_{12}(\tau_{12})$ by $G_{12}\leftrightarrow G_{21}$. 
In Appendix \ref{DiagSD} we show how to derive these equations diagrammatically in both the coupled SYK and tensor models.
 
\begin{figure}[h!]
  \begin{center}  
    \includegraphics [width=0.32\textwidth, angle=0.]{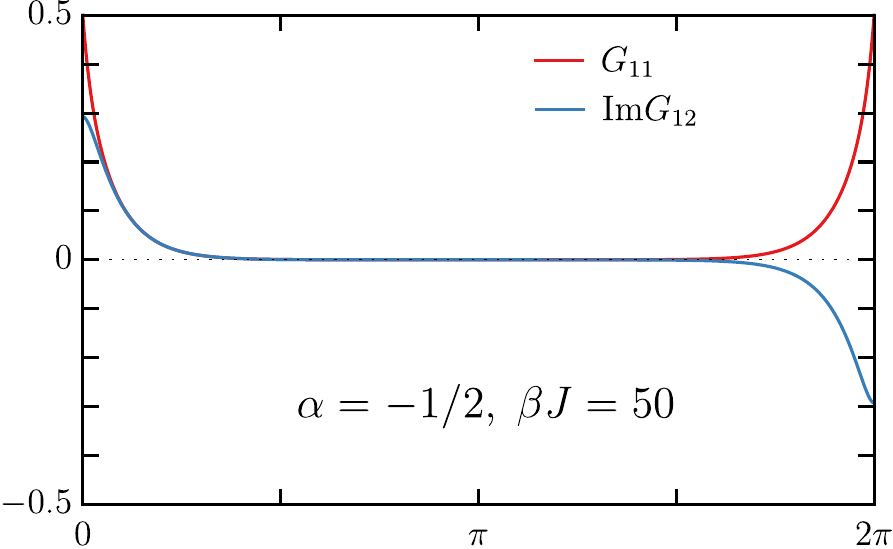}
    \includegraphics [width=0.32\textwidth, angle=0.]{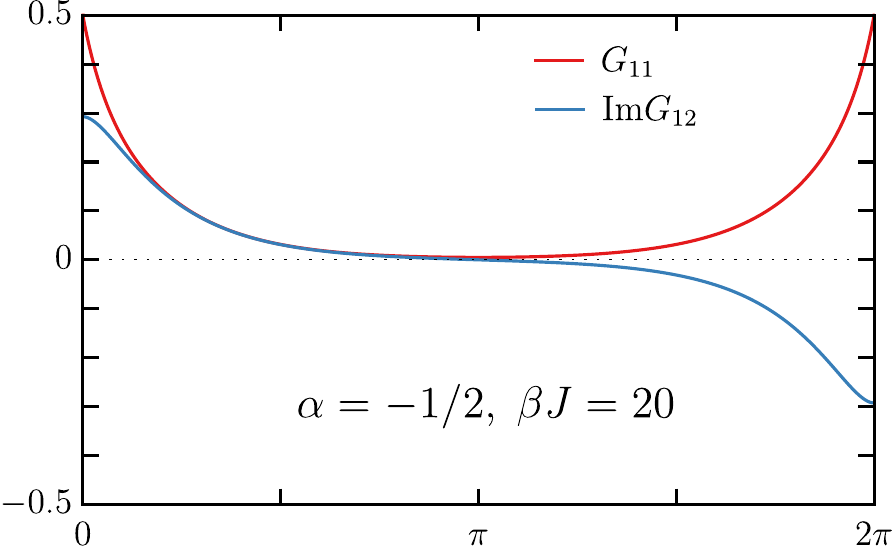}
    \includegraphics [width=0.32\textwidth, angle=0.]{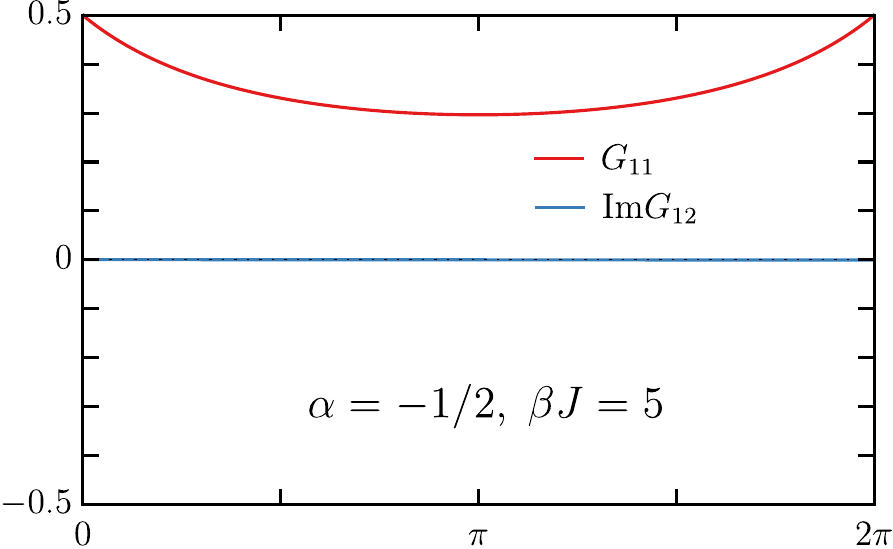}
    \includegraphics [width=0.32\textwidth, angle=0.]{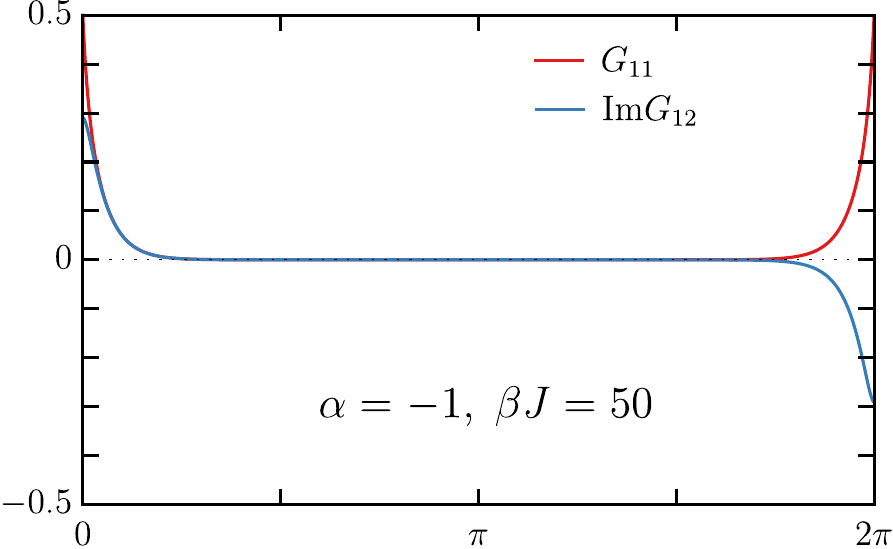}
    \includegraphics [width=0.32\textwidth, angle=0.]{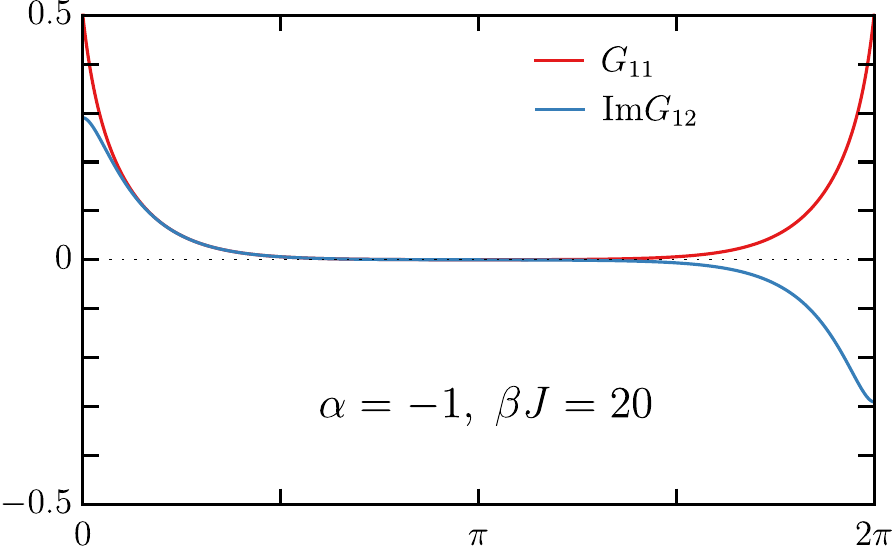}
    \includegraphics [width=0.32\textwidth, angle=0.]{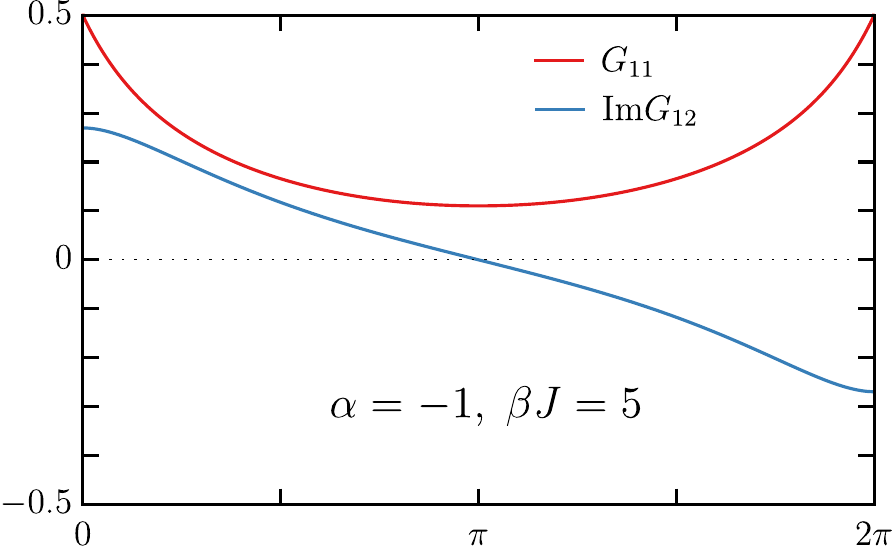}
        \includegraphics [width=0.32\textwidth, angle=0.]{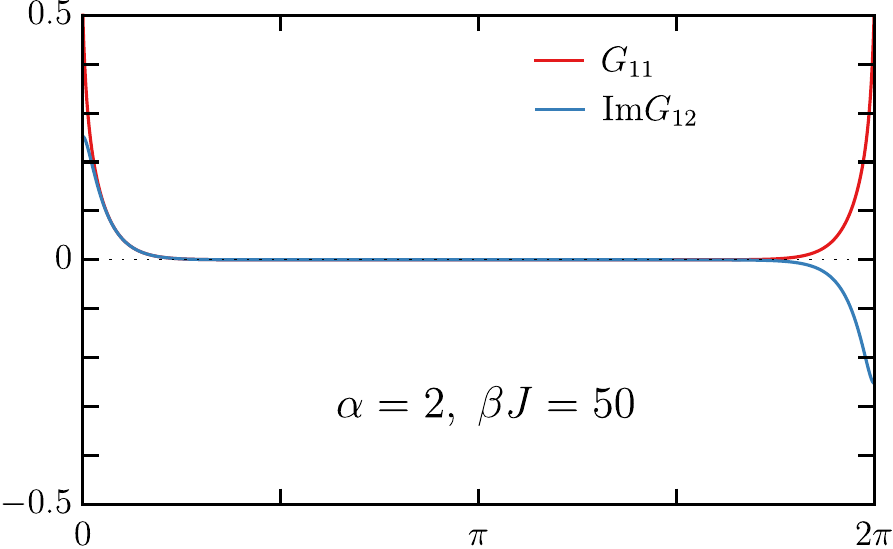}
    \includegraphics [width=0.32\textwidth, angle=0.]{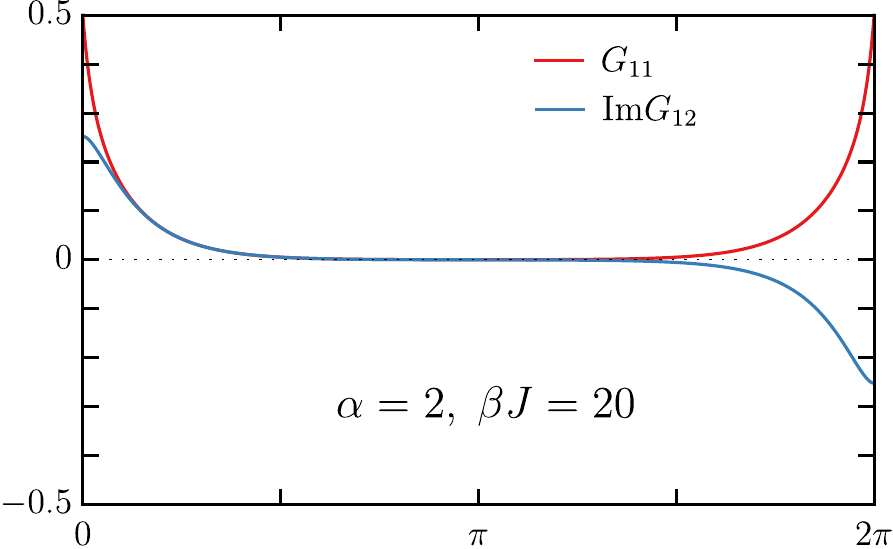}
    \includegraphics [width=0.32\textwidth, angle=0.]{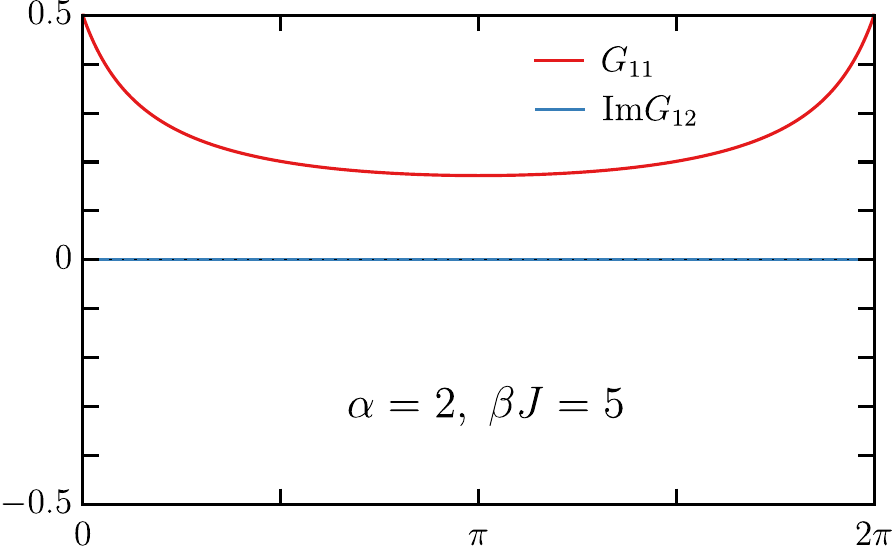}
  \end{center}
\caption{Numerical solutions to the DS equations for different values of $\alpha$ and $\beta J$, plotted against $\theta=\frac{2\pi \tau}{\beta}.$ 
All the values of $\alpha$ shown lie in the range where $U(1)_-$ is spontaneously broken at large $\beta J$.
We note that all correlators exponentially decay at the same rate.  
}
\label{Greenfunplot}
\end{figure}

 One can see that matrix $G_{\sigma\sigma'}(\tau)$ is Hermitian $G^{\dag}(\tau)=G(\tau)$, which implies that 
 $G_{11}^{*}(\tau)=G_{11}(\tau)$ and $G_{12}^{*}(\tau)=G_{21}(\tau)$. The Particle-Hole symmetry 
 implies that $G_{\sigma\sigma'}(\tau)=-G_{\sigma'\sigma}(-\tau)$, which leads to 
\begin{equation}
G_{12}(-\tau)=-G_{21}(\tau)=-G_{12}^{*}(\tau)\ .
\label{imreal}
\end{equation} 
Assuming also that $G_{22}(\tau)=G_{11}(\tau)$, we find for the DS equations 
  \begin{align}
J^{-2} \Sigma_{11}(\tau) =& (1+8\alpha^{2})G_{11}(\tau)^3  +4\alpha  G_{11}(\tau)
\big(G_{12}^{2}(\tau)+G_{12}^{*2}(\tau)+2 \alpha |G_{12}(\tau)|^{2}\big)\,,\notag\\
J^{-2} \Sigma_{12}(\tau) =& G_{12}^{3}(\tau) + 8\alpha G_{12}(\tau)G_{11}^{2}(\tau)+8 \alpha^{2}G_{12}^{*}(\tau)(G_{11}^{2}(\tau)+|G_{12}(\tau)|^{2})\ , \label{SDeq2}
\end{align} 
together with $\Sigma_{22}(\tau)=\Sigma_{11}(\tau)$ and $\Sigma_{21}(\tau)=\Sigma_{12}^{*}(\tau)$. We notice that $G_{11}(\tau)$ is real, $G_{11}^{*}(\tau)=G_{11}(\tau)$, whereas $G_{12}(\tau)$ can be complex. 
 The first series of DS equations then reads
   \begin{align}
&\partial_{\tau}G_{11}(\tau) - \int d\tau' \big(\Sigma_{11}(\tau-\tau')G_{11}(\tau')+\Sigma_{12}(\tau-\tau')G_{12}^{*}(\tau')\big) = \delta(\tau)\,, \notag\\
&\partial_{\tau}G_{12}(\tau) - \int d\tau' \big(\Sigma_{11}(\tau-\tau')G_{12}(\tau')+\Sigma_{12}(\tau-\tau')G_{11}(\tau')\big) =0\,.
\end{align} 
 
Now we can look for solutions preserving different kinds of discrete symmetries. If we assume that the solution preserves the $Z_4$ symmetry (\ref{zfour}),\footnote{
Alternatively, we may assume an interchange symmetry $c_{1i}\leftrightarrow c_{2i}$, which implies $G_{12} (\tau) =G_{21}(\tau)$.
Combining this with $G_{12}^*(\tau) = G_{21} (\tau)$,  we see that $G_{12}$
is now purely real and odd. 
Thus, we have two odd real functions: $G_{11}(\tau)$ and  $G_{12}(\tau)$. In this phase there cannot be a VEV of operator $c_{1i} c_{2i}^\dagger$, but there can be a VEV of
$c_{1i}\partial_\tau c_{2i}^\dagger$. However, the latter is unlikely to appear dynamically. Therefore, the interchange symmetry does not appear to be realized.}  
then we have $G_{12} (\tau) =- G_{21}(\tau)$. Combining this with $G_{12}^*(\tau) = G_{21} (\tau)$, we see that $G_{12}$
is purely imaginary. 
Using also (\ref{imreal}), we find that $G_{12}(\tau) = G_{12} (-\tau)$. Therefore, similarly to \cite{Kim:2019upg}, we have to solve for only two functions: an odd real one, $G_{11}(\tau)= G_{22}(\tau)$, and
an even imaginary one, $G_{12}(\tau)$. The equations determining these two functions are 
 \begin{align}
J^{-2} \Sigma_{11}(\tau) =& (1+8\alpha^{2})G_{11}^3(\tau)  +8\alpha (1-\alpha)  G_{11}(\tau) G_{12}^{2}(\tau) \,,\notag\\
J^{-2} \Sigma_{12}(\tau) =& (1+8\alpha^{2}) 
G_{12}^{3}(\tau) + 8\alpha (1-\alpha) G_{12}(\tau)G_{11}^{2}(\tau)\ .\label{SDeqcompact}
\end{align} 
They are very similar to the equations derived in \cite{Kim:2019upg}; the functions of $\alpha$ are somewhat different, but they again demonstrate changes of behavior at $\alpha=0$ and $1$.
The solutions to these equations may be obtained similarly to those in \cite{Kim:2019upg}, and they
 are plotted in fig. \ref{Greenfunplot}. 
 
 We note that there is a duality symmetry of (\ref{SDeqcompact}): these equations are invariant under 
\begin{equation}
J\to\frac{1+8\alpha}{3}J\ ,\qquad  \alpha \to \frac{1-\alpha}{1+8\alpha}\ . 
\label{fakedual}
\end{equation}
However, this is not a symmetry of the theory even in the large $N$ limit: neither (\ref{SDeq2}), nor the bilinear spectrum (\ref{spec}) respect it.

Due to the underlying $U(1)_-$ symmetry, there is a continuous family of solutions obtained from these ones through the transformation $G_{12}(\tau) \rightarrow e^{i\phi} G_{12}(\tau)$. If we don't a priori assume the $Z_4$ symmetry  (\ref{zfour}), we find that the general numerical algorithm typically converges to a solution of this form with
some phase $\phi$. We note that such a solution has a modified discrete symmetry $c_{1j}\rightarrow e^{-i \phi} c_{2j}$,  $c_{2j}\rightarrow -e^{i \phi} c_{1j}$.

Let us calculate the expectation values of the $U(1)\times U(1)$ charges. After introducing a point splitting regulator and writing 
\begin{equation}
Q_1= \lim_{\epsilon\rightarrow 0} \frac 1 2 [c^{\dag}_{1i}(\epsilon) ,  c_{1i}(0)] \ ,
\end{equation}
it follows that 
\begin{equation}
\langle Q_1\rangle=\frac{1}{2}\lim_{\epsilon\rightarrow 0^+} \left(G_{11}(\epsilon) +  G_{11}(-\epsilon)\right)= 
\frac{1}{2}\lim_{\epsilon\rightarrow 0^+}  \left(G_{11}(\epsilon) - G_{11} (\beta-\epsilon) \right ) \ .
\label{Qvev}
\end{equation}
Since for the solution in fig.\ref{Greenfunplot}  $G_{11} (\tau) =G_{22} (\tau)$ has the symmetry $G_{11} (\tau) = G_{11} (\beta-\tau)$, we see that
\begin{equation}
\langle Q_+\rangle=\lim_{\epsilon\rightarrow 0} \frac{1}{2}\left(G_{11}(\epsilon)-G_{11}(\beta-\epsilon)+G_{22}(\epsilon)-G_{22}(\beta-\epsilon) \right)=0 \ . 
\end{equation}
Analogously, we see that $\langle Q_-\rangle=0$. 
Since $U(1)_+$ is unbroken, $\langle Q_+\rangle=0$ indicates that any ground state must have $Q_+=0.$ For the broken symmetry $U(1)_-$, $\langle Q_-\rangle=0$  follows from the charge conjugation symmetry we imposed on the solution. To see this we note that, in the large $N$ limit, the ground state admits decomposition in the eigenstates of $Q_-$: 
$|0\rangle= \sum_{q_-} c_{q_-} |q_-\rangle. $ The charge conjugation symmetry implies that $c_{q_-}=c_{-q_-}$; therefore, $\langle Q_-\rangle= \sum_{q_->0} q_-(c_{q_-}-c_{-q_-})=0.  $ 

  \begin{figure}[h!]
  \begin{center}  
    \includegraphics [width=0.5\textwidth, angle=0.]{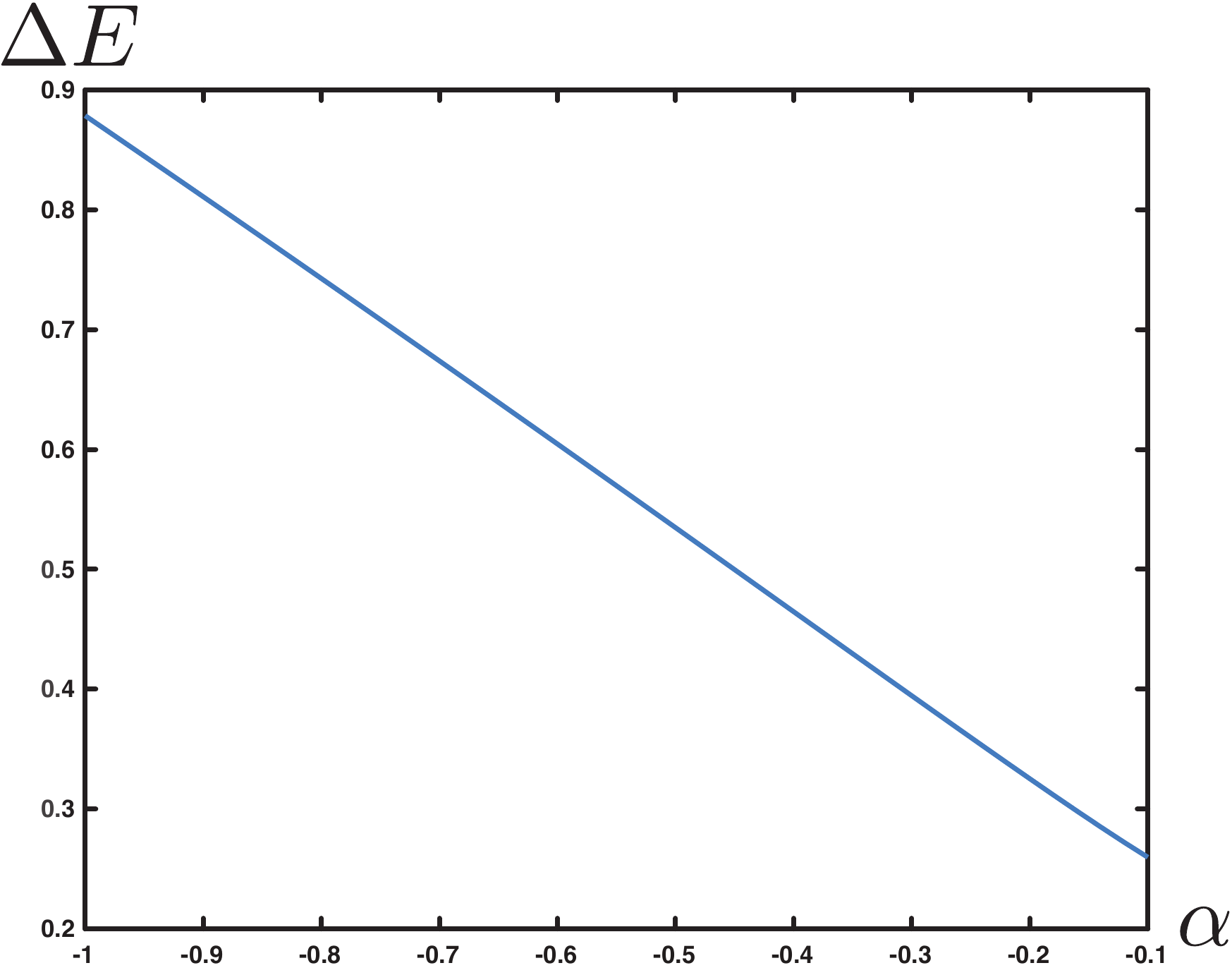}
  \end{center}
\caption{The gap $\Delta E$ (in units where $J=1$) for negative $\alpha$ calculated from the exponential 
decay of the solutions to the DS equations. We obtain the gap by linear 
fitting $\log|G_{11}|$ in regime $\frac{1}{J} \ll \tau\ll \beta,$ where the solution is dominated by the exponential decay, and the exponent is dominated by $\Delta E$ at zero temperature. 
}
\label{Gapplot}
\end{figure}

The exponential decay in fig. \ref{Greenfunplot} indicates a $\mathcal{O}(1)$ gap at large $N$ between the ground state, which has $Q_+=0$, and the state with the lowest energy in the $Q_+=1$ sector, i.e. 
\begin{equation}
\Delta E=
E_0 (Q_+=1)- E_0 (Q_+=0)\ .
\label{DSgap}
\end{equation}
 To see this, consider inserting a complete set of states 
\begin{equation}
\langle c^{\dag}_\sigma(\tau)c_{\sigma'}(0)\rangle= \sum_n e^{(E_0-E_n)\tau}\langle 0|c^{\dag}_\sigma |n\rangle \langle n| c_{\sigma'}|0\rangle\,,
\end{equation}
where $\sigma,\sigma'$ ranges from 1 to 2. In order for the matrix elements to be non-vanishing, $|n\rangle$ must have $Q_+=1. $ Using the numerical solutions to DS equations, extrapolated to large $\beta J$, we have plotted in fig. \ref{Gapplot} the quantity $\Delta E$ from (\ref{DSgap}).

  \begin{figure}[h!]
  \begin{center}  
    \includegraphics [width=0.5\textwidth, angle=0.]{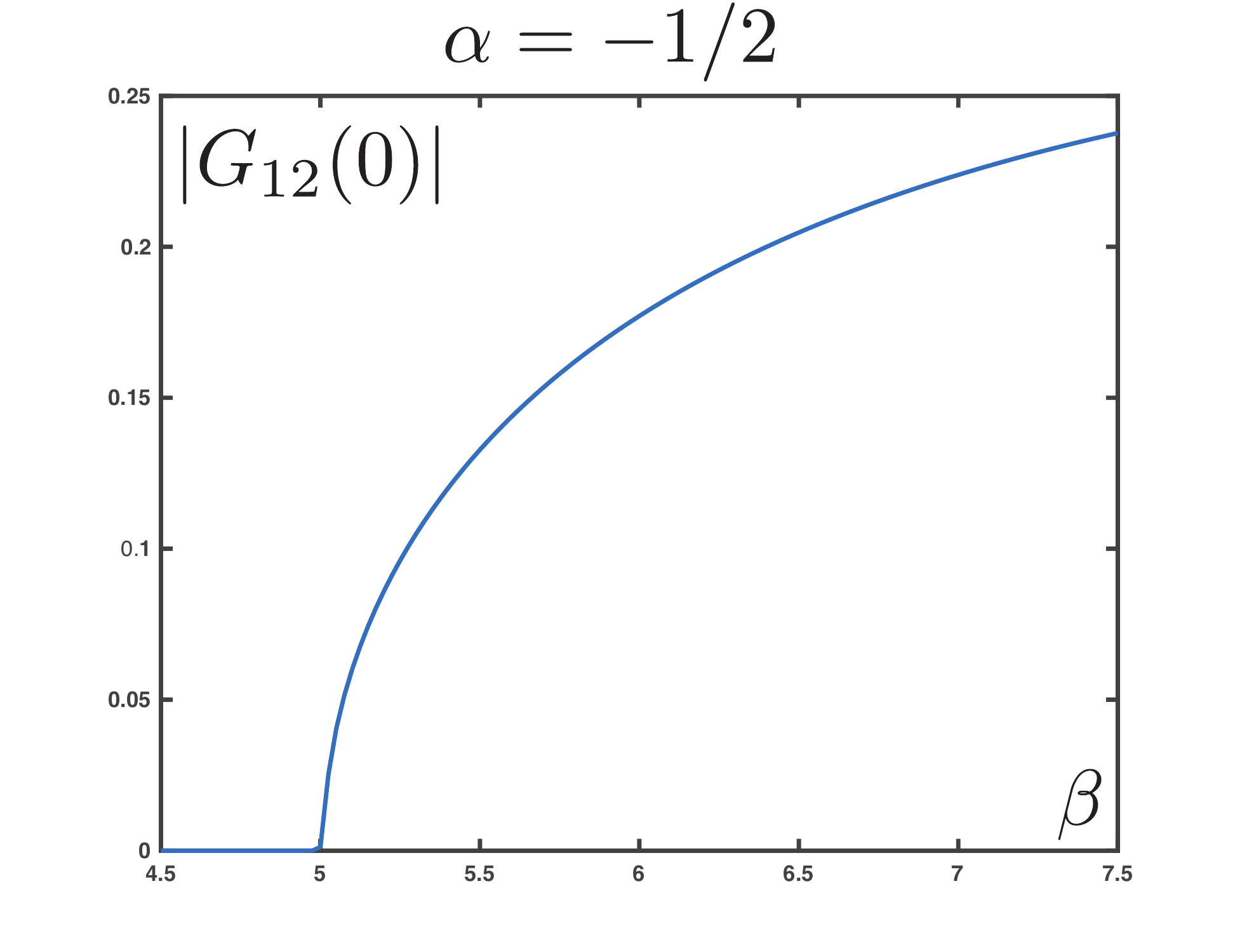}
  \end{center}
\caption{The magnitude of the off diagonal correlator at $\tau=0$, corresponding to the VEV of the operator ${c}_{1i} c_{2i}^\dagger$, plotted as a function of $\beta$ (we use units where $J=1$). 
A similar symmetry breaking behavior is observed for other values $\alpha<0$ or $\alpha>1$.    
}
\label{Freeplot}
\end{figure}

  \begin{figure}[h!]
  \begin{center}  
    \includegraphics [width=0.51\textwidth, angle=0.]{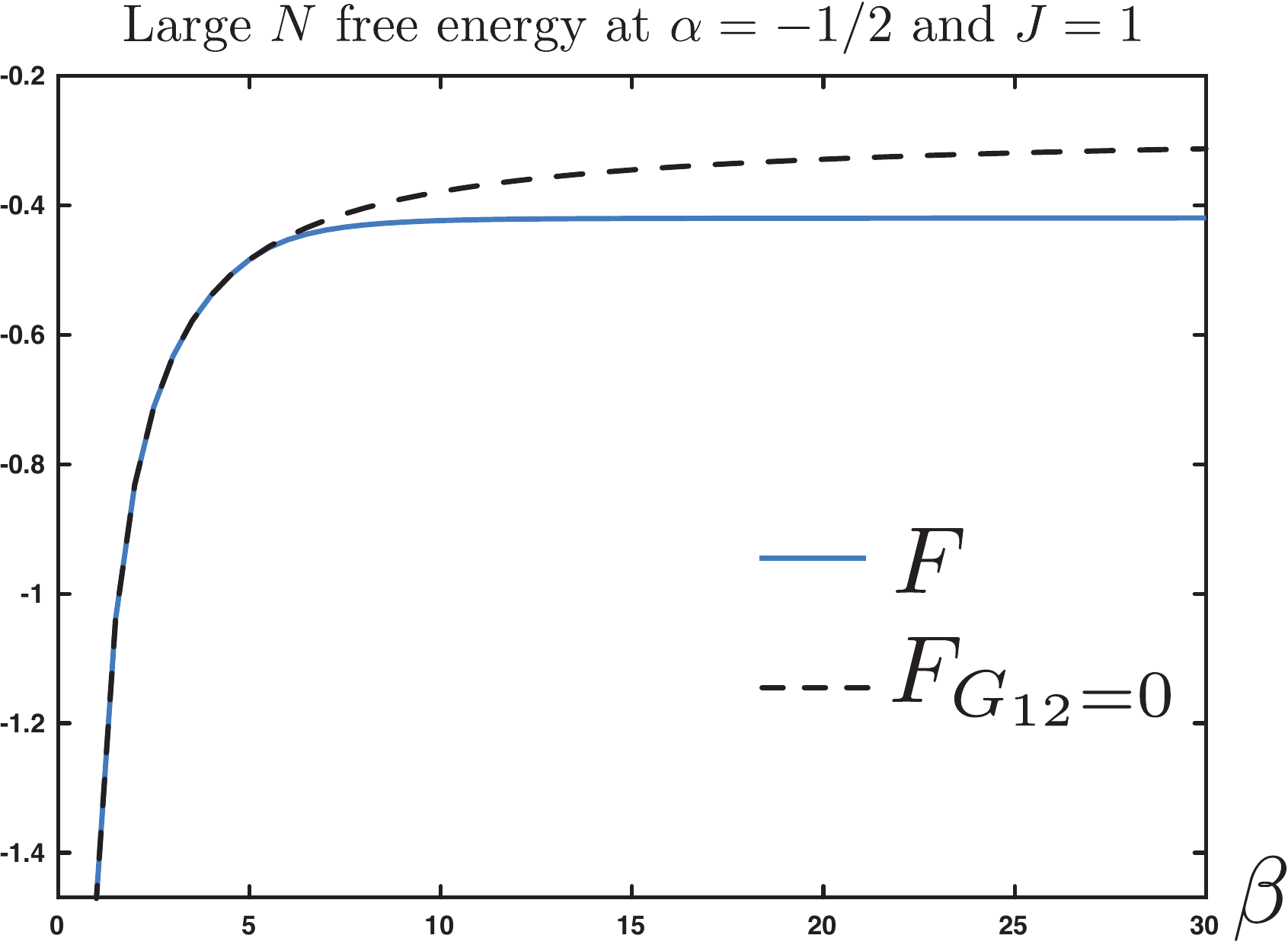}
    \includegraphics [width=0.47\textwidth, angle=0.]{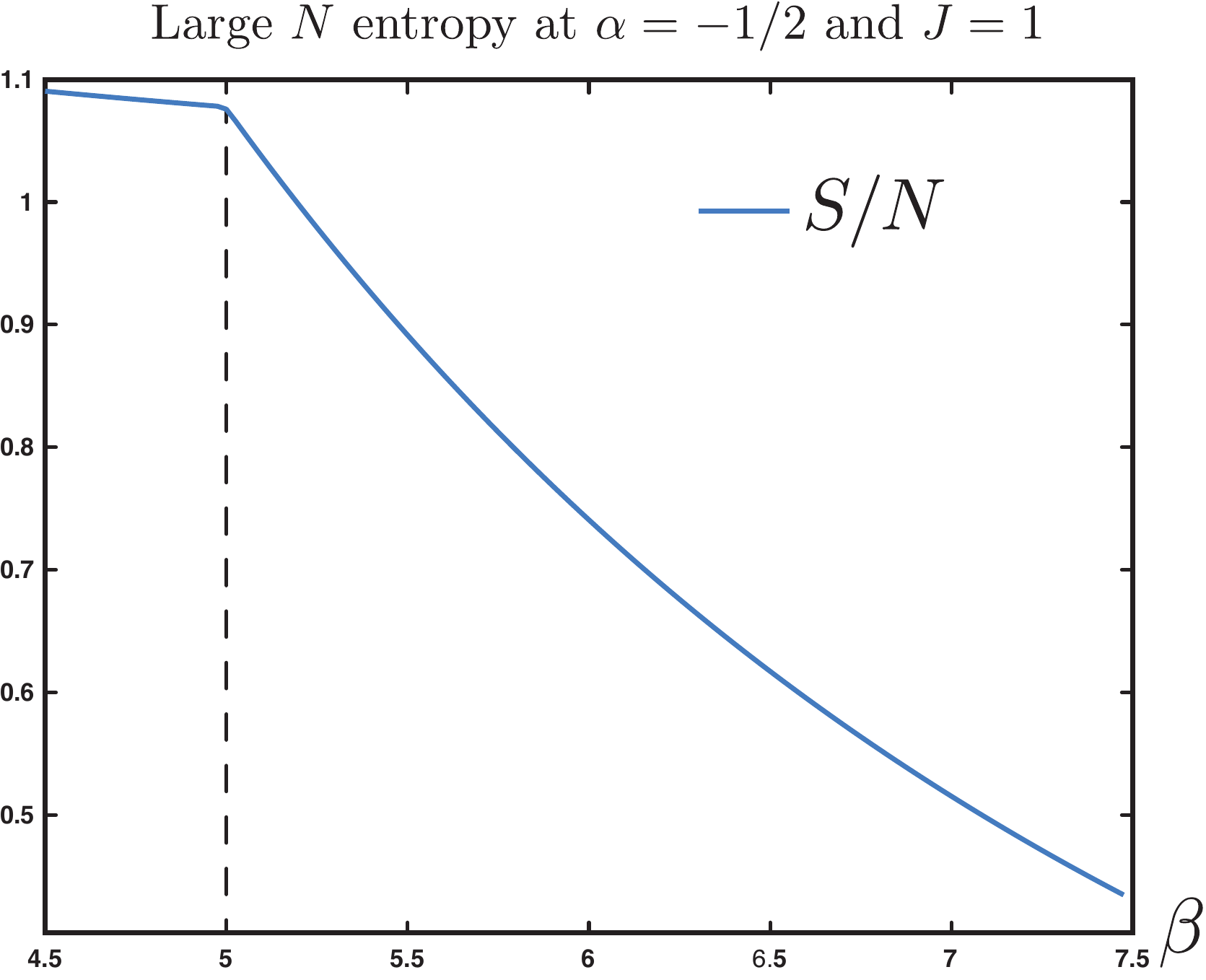}
  \end{center}
\caption{Numerical calculation of the large $N$ free energy and entropy at $\alpha=-\frac{1}{2}.$ Similarly to \cite{Kim:2019upg}, for fixed $\alpha$ we observe a second-order phase transition from the $U(1)$ symmetric phase to $U(1)$ broken phase. 
We numerically observe that $S/N$ approaches zero, rather than a finite number, as $\beta\to \infty.$ This may be explained by the $U(1)$ sigma model, where one expects $S_0 \sim \log N$ instead of powers in $N.$  
}
\label{Freeplotnew}
\end{figure}

Given the DS solution, we can also calculate the ground state energy via 
\begin{equation}
\langle 0| H |0\rangle =\lim_{\epsilon\to 0^+}\frac{1}{2}\left(\langle c^{\dagger}_{1i}(\tau+\epsilon)\partial_\tau c_{1i}(\tau)\rangle+\langle c^{\dagger}_{2i}(\tau+\epsilon)\partial_\tau c_{2i}(\tau)\rangle\right)=\lim_{\tau\to 0^+}\partial_\tau G(\tau). 
\end{equation}
In momentum space this is given by 
\begin{equation}
E_0=\frac{1}{\beta}\sum_n(\Sigma_{11}(\omega_n) G_{11}(\omega_n)-\Sigma_{12}(\omega_n) G_{12}(\omega_n))\ .
\label{EgroundDS}
\end{equation}
We find good agreement between the DS computation of the ground state energy and the exact diagonalization results, as summarized in fig. \ref{Eground}. 

\section{Charge compressibility and the sigma model}\label{compressibility}

Since the $U(1)_{-}$ symmetry is spontaneously broken for $\alpha<0$ and $\alpha>1$, we expect the presence of a gapless Goldstone mode. It arises from the degeneracy between ground states in sectors with different values of the charge $Q_-$, which emerges in the large $N$ limit.
The expected action for the Goldstone modes is the $U(1)$ sigma model action:
\beq
\label{u1_minus_sigma_model}
S_{U(1)_{-}} = \frac{N K_-}{2} \int d\tau \ (\pr_\tau \phi(\tau))^2\,, \quad  \phi \sim \phi+2\pi\ ,
\eeq
where the coefficient $K_-$, which is $\mathcal{O}(1)$ in the large $N$ limit, 
is the zero-temperature compressibility for the $U(1)_-$ charge. 

Let us emphasize that this $U(1)_-$ sigma model has a completely different origin from the $U(1)$ sigma model
arising in the complex SYK model, which was recently discussed in detail in
\cite{Gu:2019jub}. In the latter case, the physics is similar to the 
conventional SYK model: there is an approximate conformal symmetry in the IR, 
the Schwartzian effective action, zero-temperature entropy and, most importantly,
\textit{U(1) symmetry is not broken.} The $U(1)$ effective action for the complex SYK model \cite{Gu:2019jub} has the same origin as the Schwartzian action,
since dropping the fermionic kinetic term promotes the global $U(1)$ symmetry to local $U(1)$. 
The finite $1/J$ corrections manifest themselves in the time-reparametrization Schwartzian mode and the $U(1)$-phase
reparametrization sigma-model. 

\textit{Assuming} that in the range $0 < \al < 1$ the solution is given by the standard near-conformal
SYK saddle, so there are no anomalous
VEVs, we essentially have two non-interacting complex fermions. In particular, when chemical potential $\mu_+\neq 0$, the system has solution $G_{11}(\tau)=G_{22}(\tau), $ and when $\mu_-\neq 0,$ $G_{11}(\tau)=G_{22}(\beta-\tau ).$ Both reduce (\ref{firstsdeq}) and (\ref{secondsdeq}) to that of two decoupled complex fermions with chemical potentials $\mu_{\pm}$. 
We then find that we have 
two sigma-models, for $U(1)_\pm$, with compressibilities:
\beq
K_- = K_+ = \frac{2K_{\textrm{cSYK}}}{\sqrt{1+8 \al^2}}\ , 
\label{compal}
\eeq
where $K_{cSYK}\approx 1.04$ is the compressibility of a single complex SYK model \cite{Gu:2019jub}. The factor of two comes from having two fermions, 
and the square root comes from renormalization of $J$ by non-zero $\alpha$, (\ref{sym:J}).
Let us point out though that, at $\alpha=1/4$, the $U(1)_-$ symmetry is enhanced to $SU(2)$. 
We will discuss this case separately in section \ref{sec:su2}.

In the case of spontaneously broken $U(1)_-$ symmetry, the physics is different. The solutions of the Dyson-Schwinger equations that we have found for $\alpha<0$ do not have a conformal form. Therefore,
there is no approximate reparametrization symmetry or Schwartzian effective action.
In the large $N$ limit, the action (\ref{u1_minus_sigma_model}) is a conventional Nambu-Goldstone mode action. 
On these grounds, we do not expect to have a sigma model for $U(1)_+$ symmetry, 
since it is unbroken. Therefore, the splittings between sectors with different values of $Q_+$ 
should not vanish in the large $N$ limit. This implies that the compressibility $K_+$ defined 
as $dQ_+/d\mu_+$ is zero, so that a small chemical potential does not generate non-zero charge. We will
see this in the large $N$ DS equations momentarily. 
In the exact diagonalization at finite $N$, this manifests in the fact that 
the energy dependence on $Q_+$ is not close to quadratic.

Let us return to the $U(1)_-$ symmetry and compute the corresponding compressibility $K_-$. It can be found in three ways:
First of all, it is the derivative of the charge with respect to the chemical potential:
\beq
K_- = \frac{d Q_-}{d \mu_-}\,, \quad  \text{at} \ T=0\,, \mu_-=0\,.
\eeq
Secondly, it is related to the grand canonical thermodynamical potential $\Omega$ as 
\beq
\label{Omega_K_minus}
\Omega = \Omega_0 - \frac{N K_- \mu_-^2}{2}\,, \quad T=0\,,
\eeq
and finally the action (\ref{u1_minus_sigma_model}) can be quantized leading to the spectrum:
\beq
\label{u1_minus_spectrum}
E_{Q_-} = A+ \frac{Q_-^2}{2 N K_-}\,, \quad Q_- \in \mathbb{Z}\,.
\eeq
Let us emphasize that the $U(1)_-$ symmetry breaking occurs only in the limit $N\rightarrow \infty$. In systems with
finite numbers of degrees of freedom this does not happen. From the above spectrum we see how it happens:
if $N=\infty$ we have a classical particle on a circle (\ref{u1_minus_sigma_model}) with an  
infinite number of classical vacua. However,
finite $N$ effects quantize the action, leading to a unique ground state and spectrum (\ref{u1_minus_spectrum}).

\begin{figure}[!ht]
\centering
\includegraphics[scale=0.9]{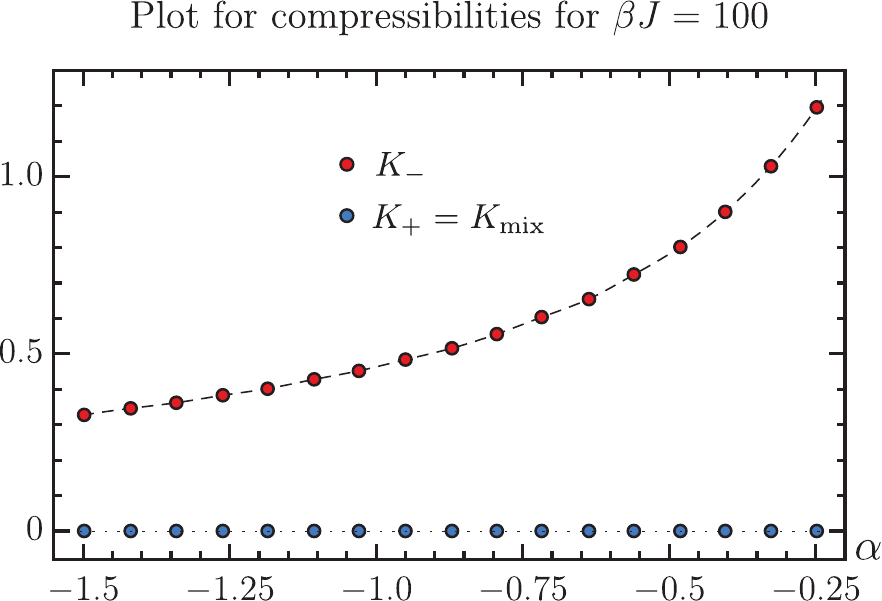}
\caption{The numerical results for three different compressibilities as functions of $\alpha$ for 
$J=1$, $\beta=100$.
 We checked that the result does not 
depend appreciably on $\beta$ by comparing with the $\beta=50$ data. }
\label{fig:Ks}
\end{figure}

It will be convenient for us to find $K_-$ numerically by introducing a chemical potential into the large $N$
Dyson-Schwinger equations and fitting
the numerical result for $\Omega$ using eq. (\ref{Omega_K_minus}). In fact, to 
double check our results, we will
introduce chemical potentials $\mu_{-}$ and $\mu_+$ for $U(1)_-$ and $U(1)_+$ and fit $\Omega$ with
\beq
\Omega = \Omega_0 - \frac{N K_- \mu^2_-}{2} - \frac{NK_+ \mu_+^2}{2} - N K_{\rm mix} \mu_- \mu_+ \,.
\eeq
Since $U(1)_+$ is unbroken, we expect that $K_+ = K_\text{mix} = 0 $. In other words, the low energy states 
are not charged under $U(1)_+$, and the gap to states with non-vanishing $U(1)_+$ charges is big.
The result is presented in Figure \ref{fig:Ks}. We indeed see that $K_+ = K_{\rm mix} = 0 $.

\begin{figure}[!ht]
\centering
\includegraphics[scale=0.8]{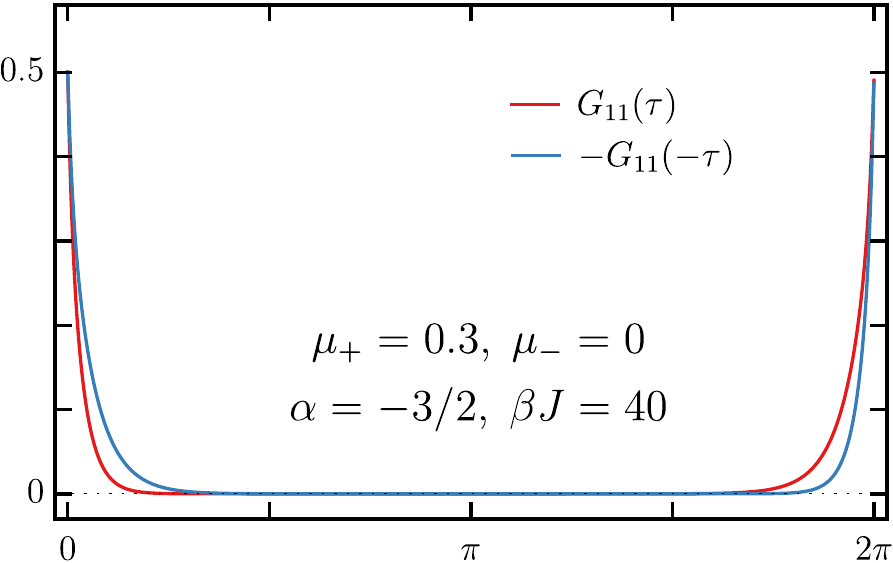}\qquad 
\includegraphics[scale=0.8]{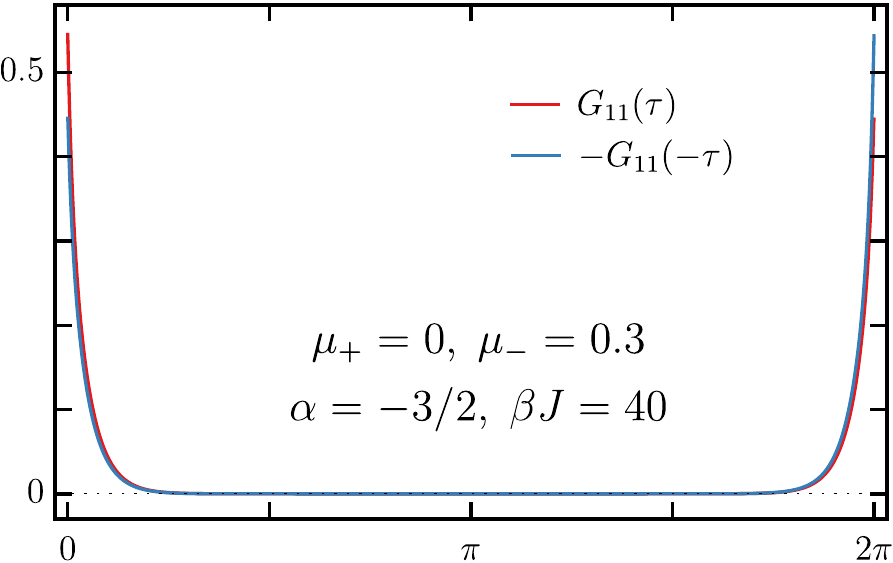}
\caption{Plots of the Green functions when chemical potentials are turned on. For $\mu_+\neq 0$, $G_{22}(\tau)= G_{11}(\tau)$, while
for $\mu_-\neq 0$, $G_{22}(\tau)= -G_{11}(-\tau)$. It follows that
$\langle Q_-\rangle $ is generated for $\mu_- \neq 0$.}
\label{fig:mus}
\end{figure}

Finally, we illustrate our claims by plotting the Green function $G_{11}$ 
upon introducing $\mu_{\pm}$  in fig. \ref{fig:mus}. 
These results were obtained by solving the DS equations numerically with $J=1, \beta=40, \alpha=-1.5$ and
 $\mu_\pm =0.3$. The expectation value of the charge $Q_1$ may be read off from the plot of $G_{11}$ using
(\ref{Qvev}), and the expectation value of $Q_2$ is analogously determined by $G_{22}$. 

From fig. \ref{fig:mus} we observe that, when $\mu_+$ is turned on, the
expectation values of $Q_1$ and $Q_2$ vanish despite the fact that the Green
functions become asymmetric around $\pi$. This means that the expectation
values of $Q_\pm = Q_1 \pm Q_2$ vanish as well.  On the other hand, when $\mu_-$ is
turned on, the asymmetry in the values of $G_{11}$ at $0+$ and $2\pi-$ is
clearly seen; the $G_{22}$ has the opposite asymmetry. 
Therefore, we now find that $\langle Q_1\rangle =-   \langle Q_2\rangle \neq 0$, so that   $\langle Q_-\rangle$ is non-vanishing.

\section{Results from Exact Diagonalizations}\label{EDsection}

In this section we will study the energy spectra for accessible values of $N$. We will use the particle-hole symmetric version of the Hamiltonian, given in 
(\ref{pholesymmetric}). 
We have generated multiple random samples of the Hamiltonain, which allow us to study various averaged quantities as functions of $\alpha$ and $N$.

\subsection{Evidence for Symmetry Breaking}

For $\alpha<0$ and $\alpha>1$, the large $N$ DS equations indicate that $U(1)_-$ symmetry is spontaneously broken. 
In these ranges of $\alpha$, the absolute ground state appears in the sectors with $Q_+=0$ and the lowest possible value of $|Q_-|$, which is  $|Q_-|=0$ for even $N$ and
$|Q_-|=1$ for odd $N$. This means that, for odd $N$, there are two degenerate ground states, which have $Q_-=\pm 1$, and their mixture admits an expectation value of operator
$c_{1i}^\dagger c_{2i}$ already at finite $N$.
At any finite even 
$N$ we cannot see the spontaneous symmetry breaking, but it appears in the large $N$ limit due to the degeneracy of ground states with $Q_+=0$ and different values of $Q_-$.

\begin{figure}[h!]

    \includegraphics [width=0.45\textwidth, angle=0.]{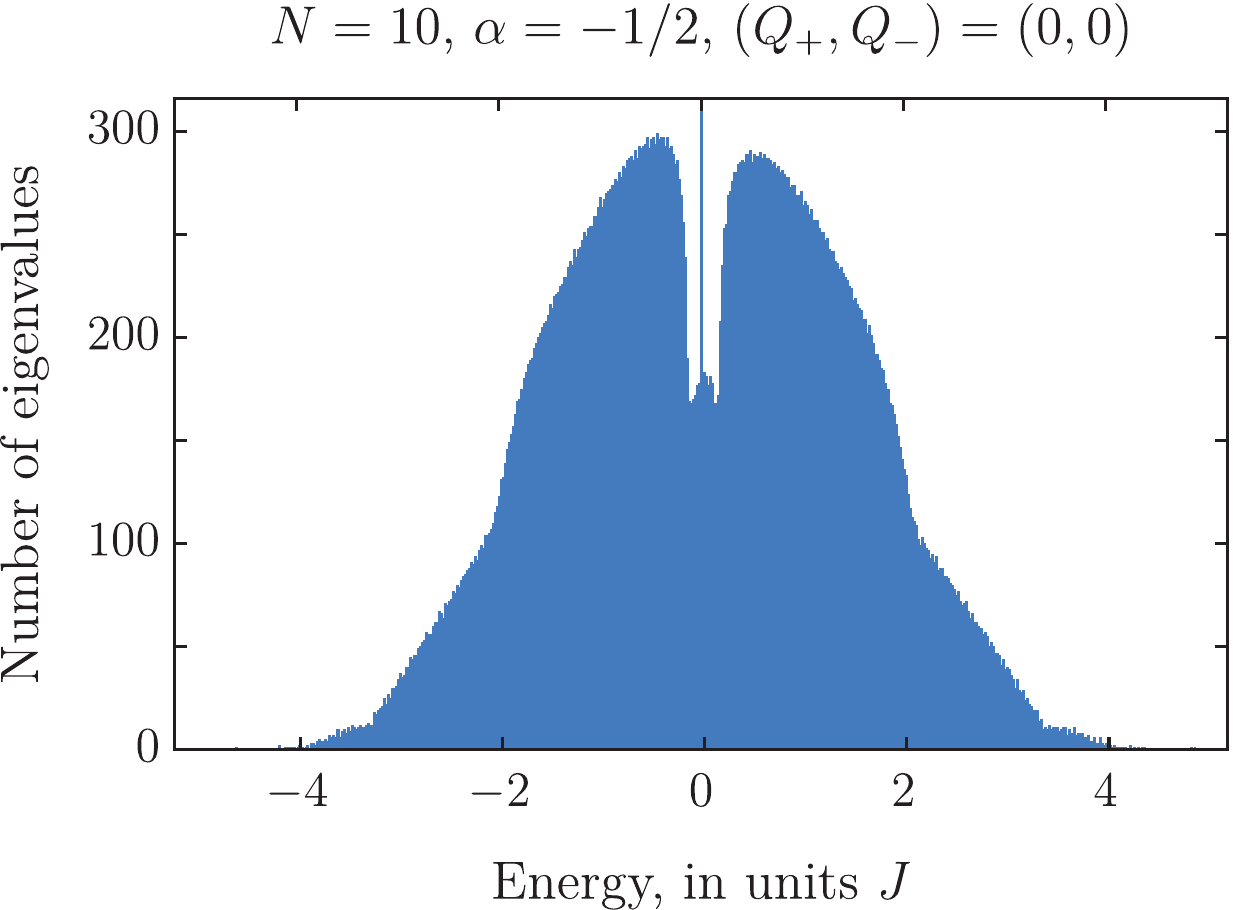} \qquad
    \includegraphics [width=0.45\textwidth, angle=0.]{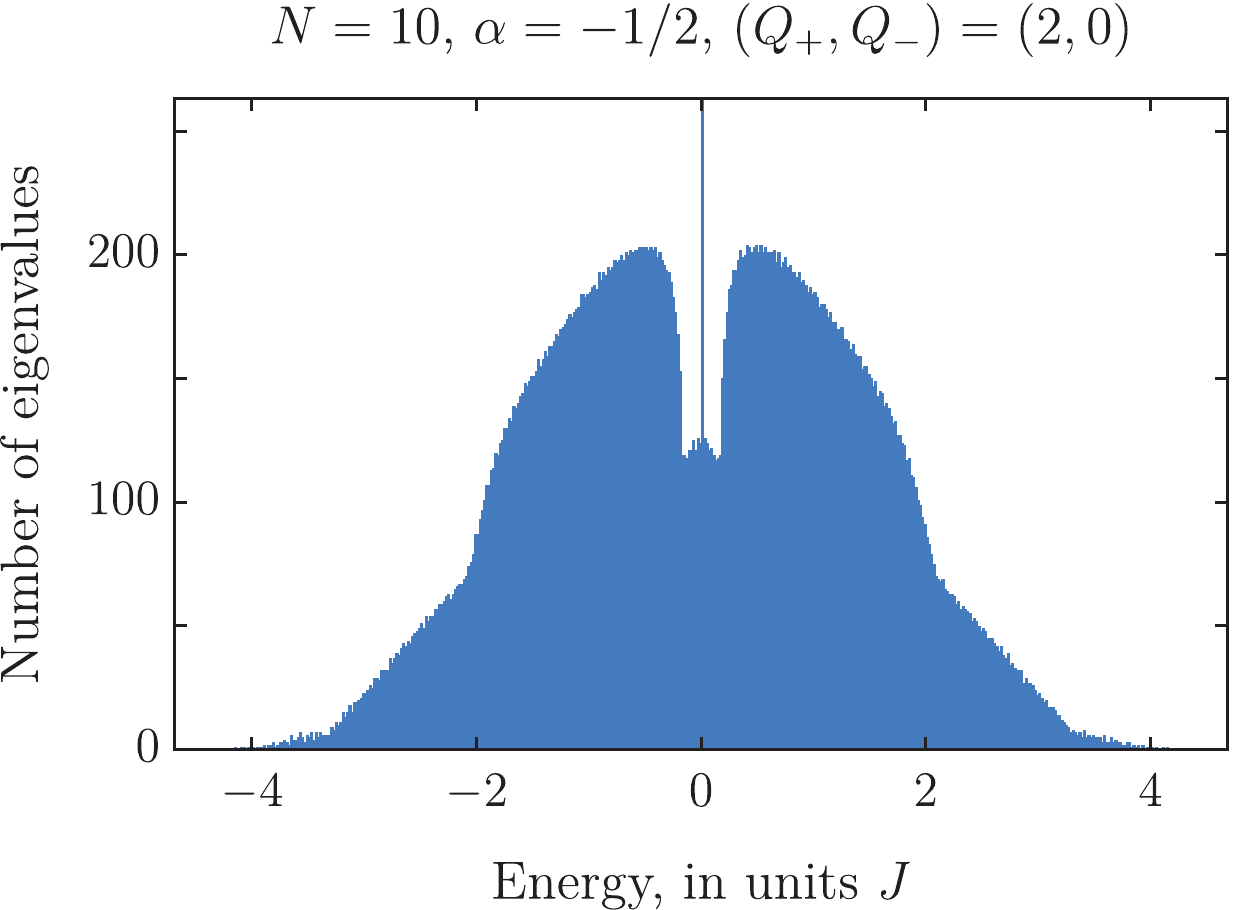} 
   \\
   \\
        \includegraphics [width=0.45\textwidth, angle=0.]{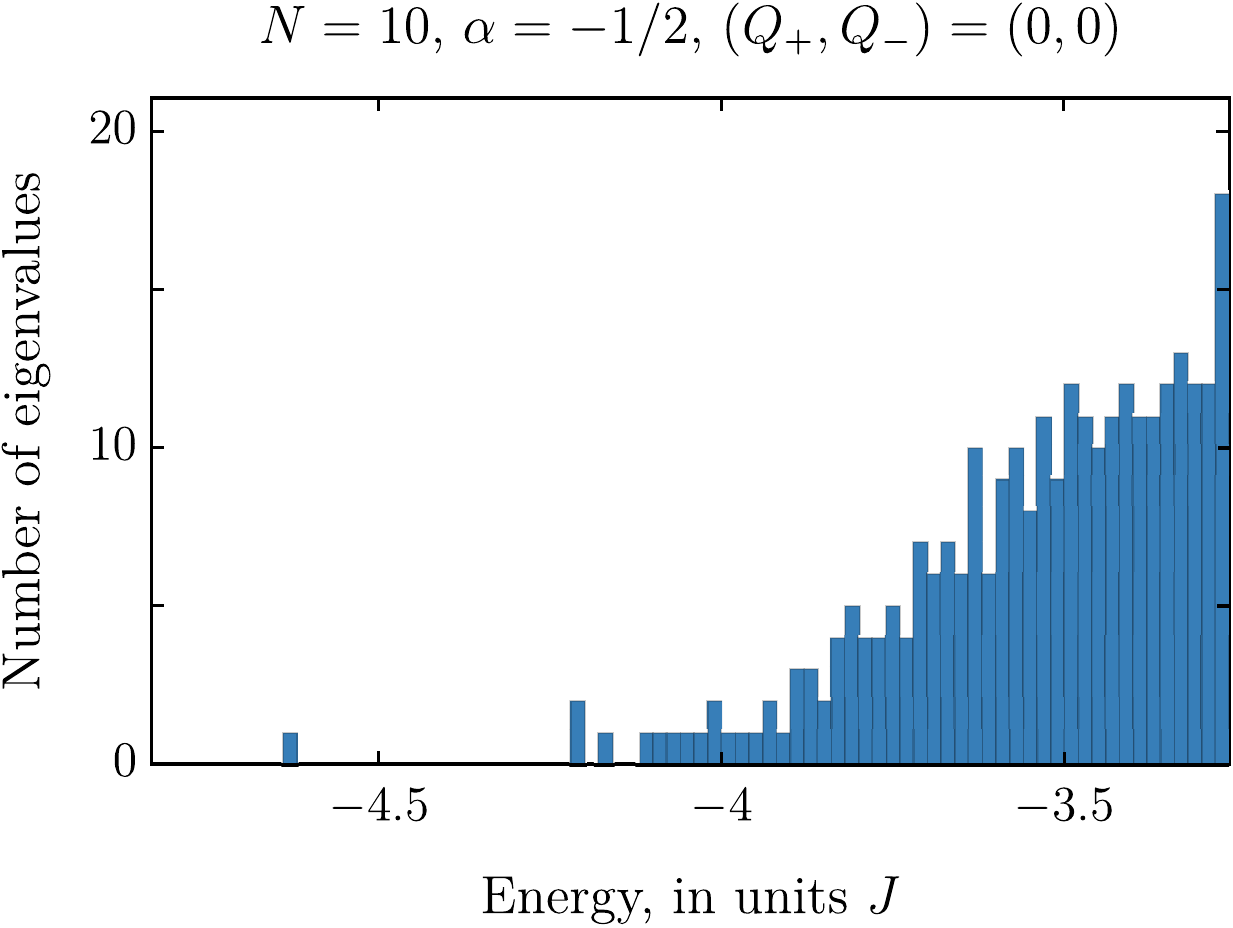} \qquad 
    \includegraphics [width=0.45\textwidth, angle=0.]{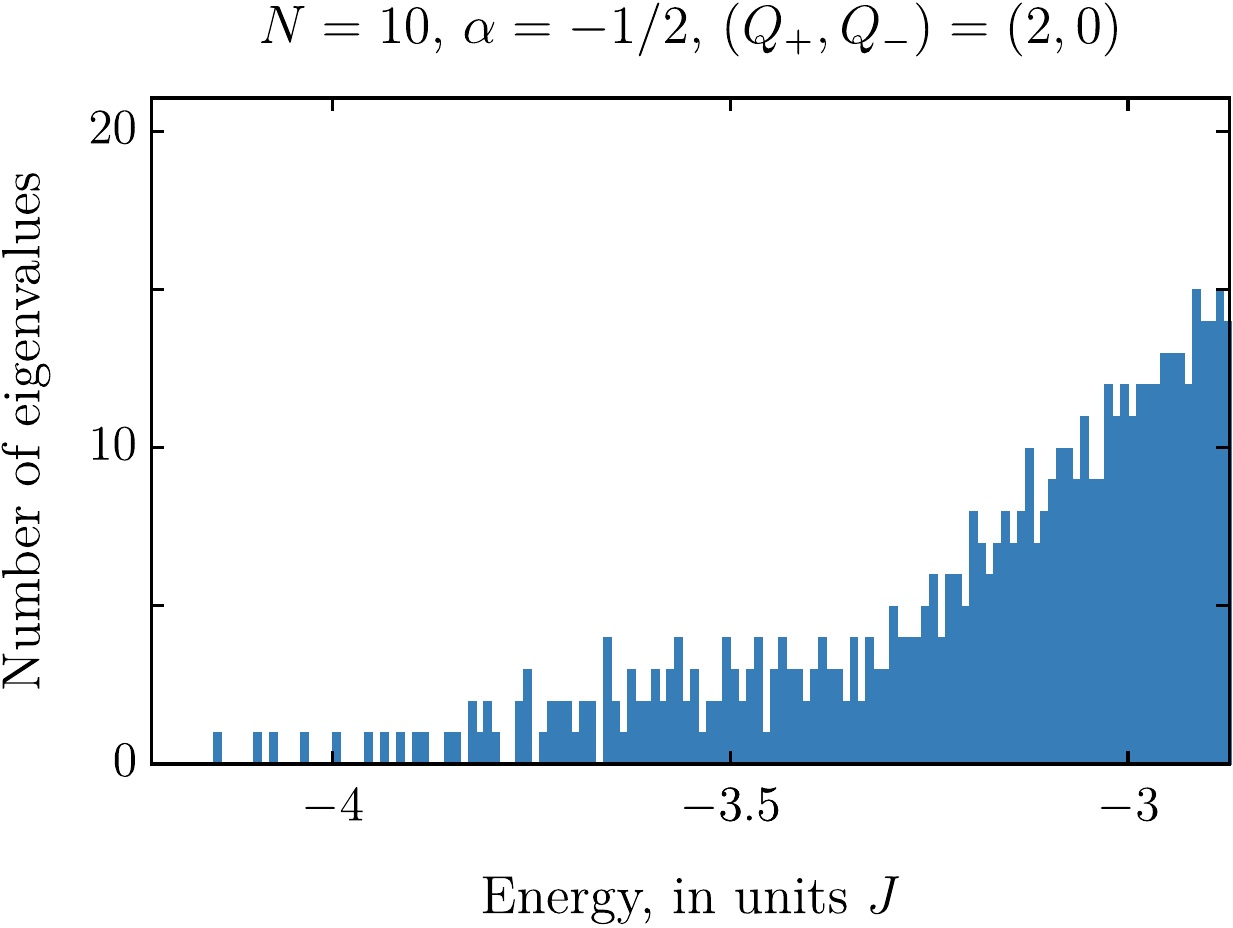}
\caption{ 
Density of states in two of the charge sectors, $(Q_{+},Q_{-})=(0,0)$ and $(2,0)$, for a single realization of the model with $N=10$ and $\alpha=-1/2$.  The lower plots are zoomed in regions near the ground state. 
We observe a prominent gap 
in the $(Q_{+},Q_{-})=(0,0)$ sector. 
}
\label{edngalpha}
\end{figure}

In fig. \ref{edngalpha} we exhibit the spectra in two different charge sectors for $N=10$. \footnote{For the special value $\alpha=-1/2$ some of the charge sectors contain a large number of states with exactly zero energy, and this number is independent of the sampling of $J_{ij,kl}$. An analogous phenomenon was observed in \cite{Kim:2019upg} for 
the coupled Majorana model with $\alpha=-1$.}  
A characteristic quantity in the broken symmetry phase is the gap between the first excited state and the ground state:
such a gap is observed in the sectors with $Q_+=0$. For example, in the $(Q_+, Q_-)=(0,1)$ sectors we find for $\alpha=-1/2$ that the average gaps above the ground state are $\approx 0.440, 0.437, 0.473$
for $N=7, 9, 11$, respectively.
These results suggest that the gap is non-vanishing in the large $N$ limit.

Similarly, there is a sizable difference between the ground state energies in sectors with different values of $Q_+$. It is noticeably bigger than 
the difference between sectors with different values of $Q_-$, which is expected to be of order $1/N$. For example, for $\alpha=-1/2$ and $N=10$, we find
\begin{gather}
E_0 (Q_+=2, Q_-=0) - E_0 (Q_+=0, Q_-=0) \approx  0.622\ , \notag \\ 
E_0 (Q_+=0, Q_-=2) - E_0 (Q_+=0, Q_-=0) \approx 0.236\ .
\end{gather}

\begin{figure}[h!]
  \begin{center}  
    \includegraphics [width=0.45\textwidth, angle=0.]{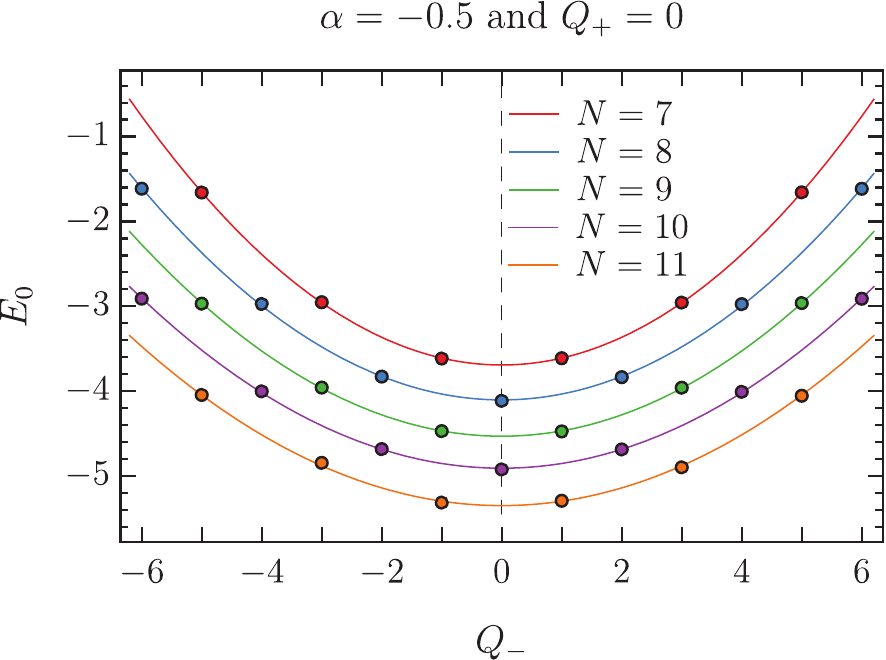}\qquad 
    \includegraphics [width=0.45\textwidth, angle=0.]{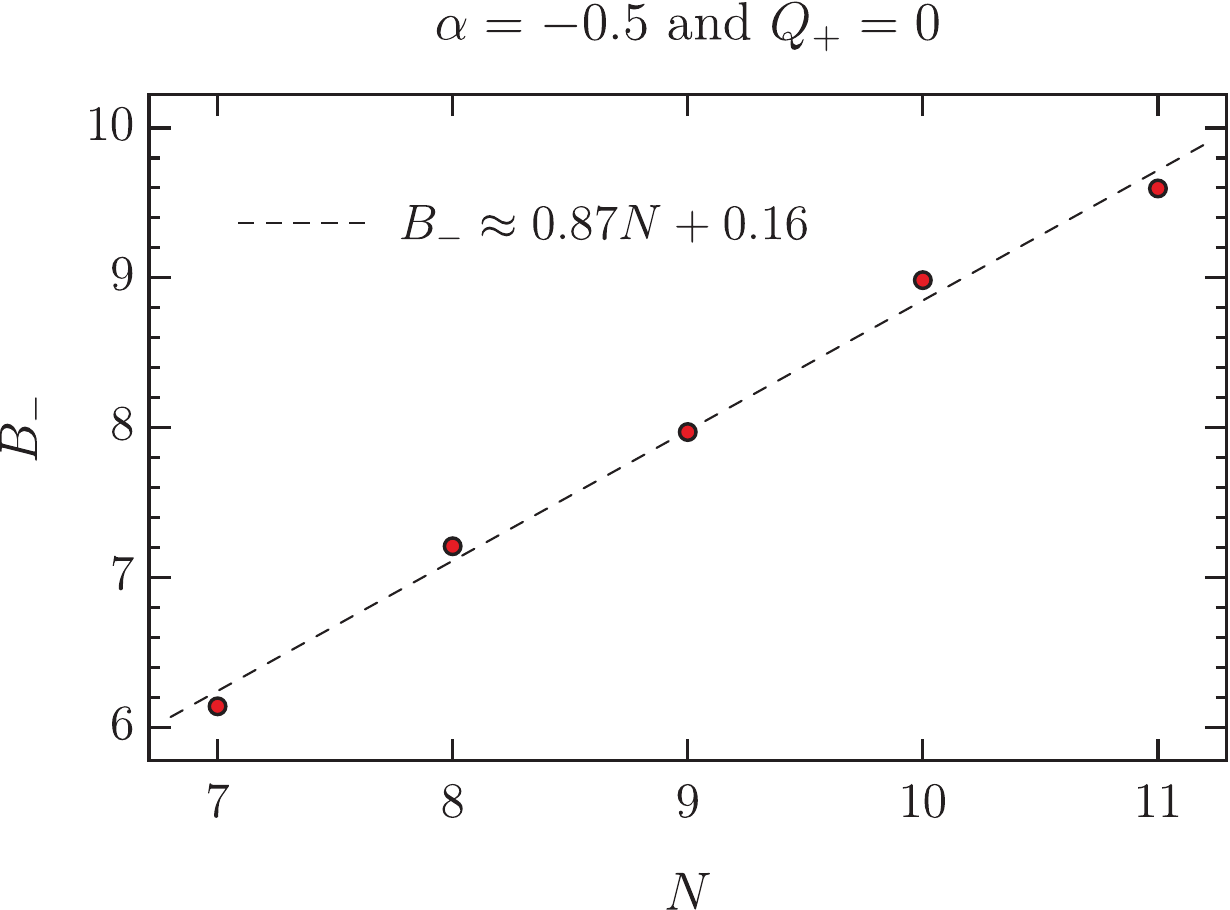}
  \end{center}  
\caption{ The $E(Q)$ curve(left) at $\alpha=-1/2$ for the first few $Q_-$ sectors at $Q_+=0.$ 
For each $E(Q)$ curve, we make a quadratic fit $E_0= \frac{Q_-^2}{2B_-}+ A$ and determine $K_-$ from the linear fit of $B_-$ vs. $N$. The slope of the plot for $\alpha=-1/2$ is  $K_{-}^{\textrm{ED}}\approx 0.87$, which agrees well with the DS calculation of $K_{-}^{\textrm{DS}}\approx 0.80$. }
\label{EQcurvemp5}
\end{figure}

\begin{figure}[h!]
  \begin{center}  
        \includegraphics [width=0.465\textwidth, angle=0.]{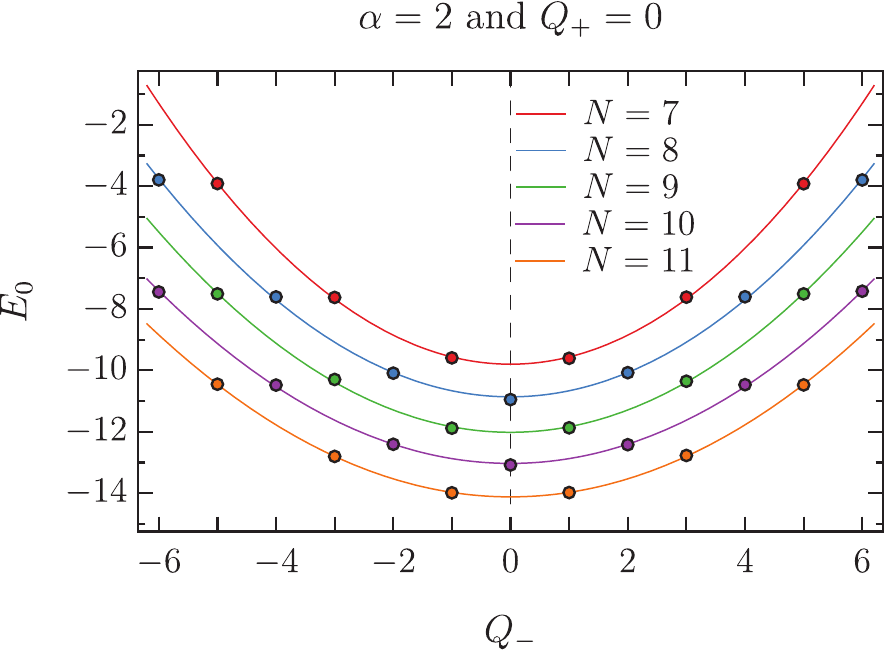}\qquad
    \includegraphics [width=0.46\textwidth, angle=0.]{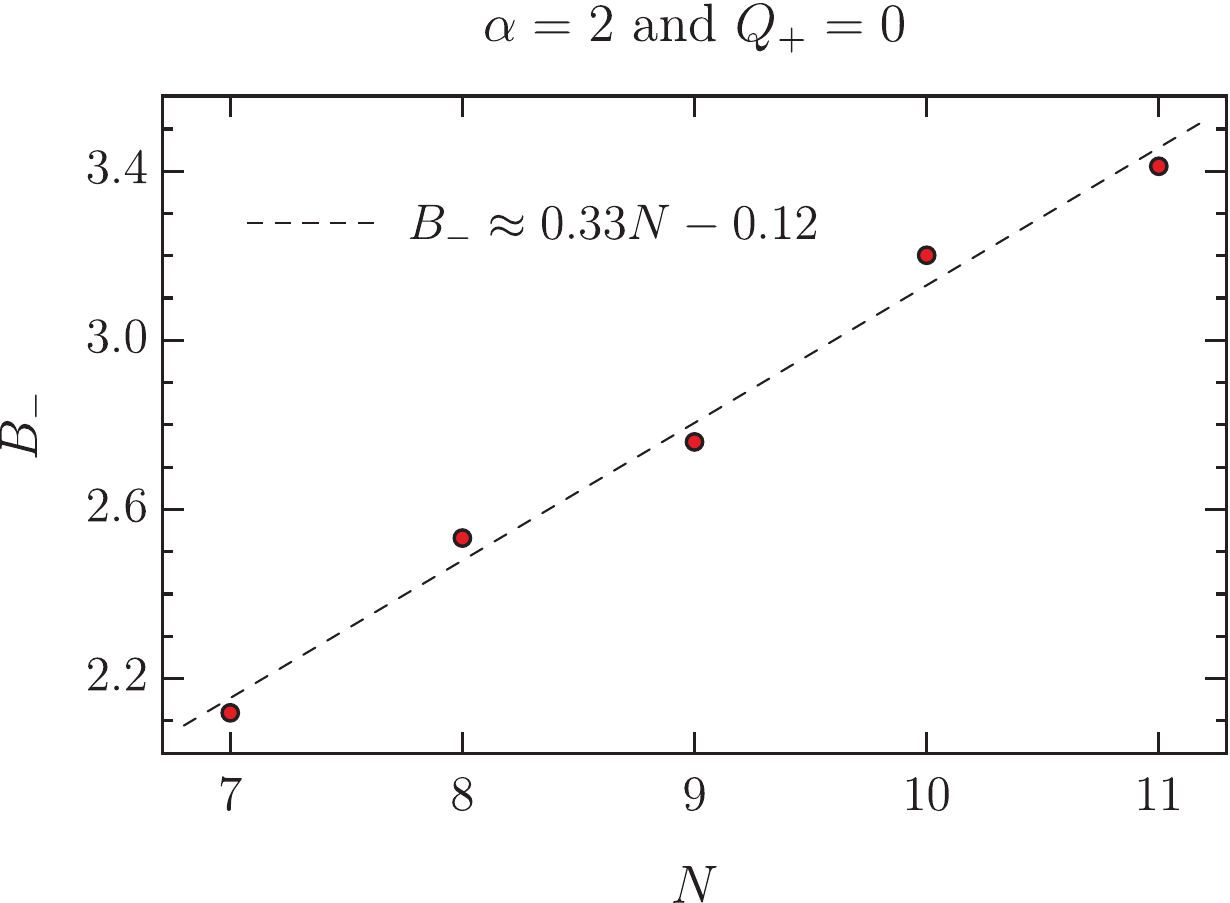}
  \end{center}  
\caption{ The $E(Q)$ curve(left) at $\alpha=2$ for the first few $Q_-$ sectors at $Q_+=0.$ 
The slope of the plot of $B_-$ vs. $N$ is  $K_{-}^{\textrm{ED}}\approx 0.33$, which is not far from the DS calculation of $K_{-}^{\textrm{DS}}\approx 0.32$. }
\label{EQcurve2}
\end{figure}

In the sectors with $Q_+=0$, we expect the ground state energies to depend quadratically on $Q_-$:
\begin{equation}
E_0= A(\alpha, N)+ 
\frac{Q_-^2}{2 B_- (\alpha, N)}\ ,  \qquad B_-(\alpha, N)= K_-(\alpha) N + C_-(\alpha) + O(1/N) \ ,
\label{Energygs}
\end{equation}
where $K_-$ is the large $N$ compressibility for the $U(1)_-$ degree of freedom.
 As can be seen in figs. \ref{EQcurvemp5} and \ref{EQcurve2}, these quadratic fits work well, and $B_-(\alpha)$ is approximately linear in $N$. 

\noindent From the slopes we find that 
$K^{\textrm{ED}}_- (-0.5) \approx 0.87$ and $K^{\textrm{ED}}_- (2) \approx 0.33$.
These values of compressibility are close to those obtained from the Dyson-Schwinger calculations directly in the large $N$ limit:  
$K^{\textrm{DS}}_- (-0.5) \approx 0.80$ and $K^{\textrm{DS}}_-(2)\approx 0.32$.

\begin{figure}[h!]
  \begin{center}  
    \includegraphics [width=0.45\textwidth, angle=0.]{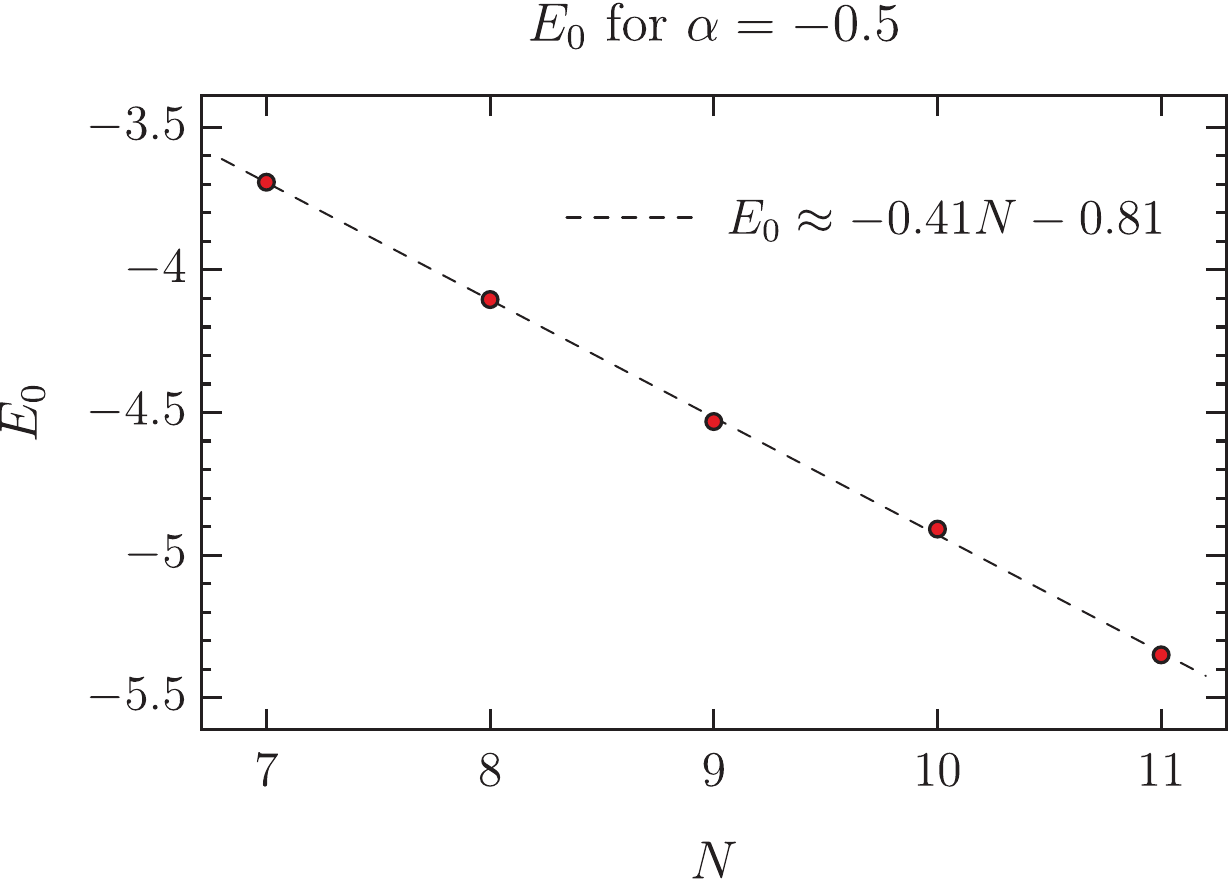}\qquad
    \includegraphics [width=0.45\textwidth, angle=0.]{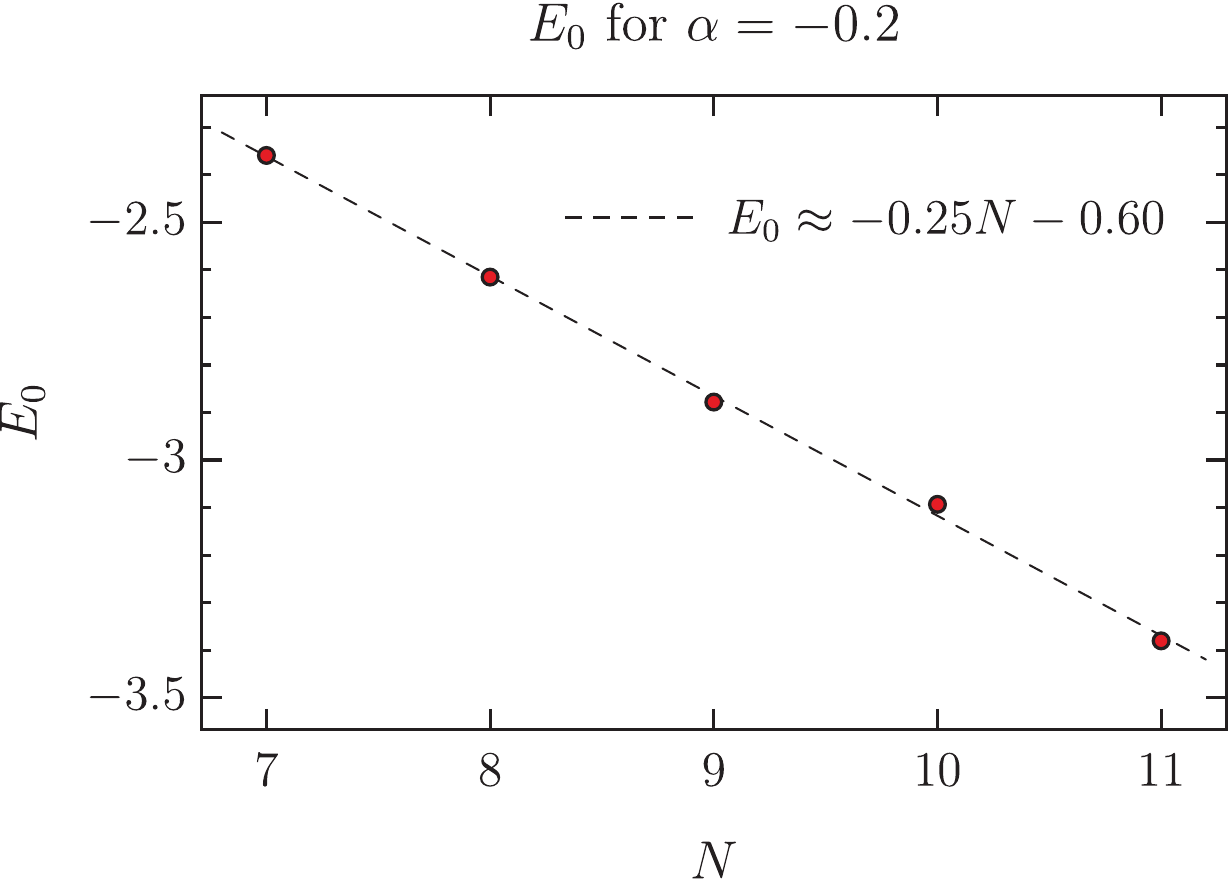}
  \end{center}  
\caption{ 
Plots of the leading term in the ground state energy (\ref{Energygs}), $A(\alpha, N)$, vs. $N$ for $\alpha=-0.5$ (left) and $\alpha=-0.2$ (right). 
The linear fits determining the slope, ${\cal E}_0 (\alpha)$, are also shown.
}
\label{Eground}
\end{figure}

\begin{figure}[h!]
  \begin{center}  
    \includegraphics [width=0.48\textwidth, angle=0.]{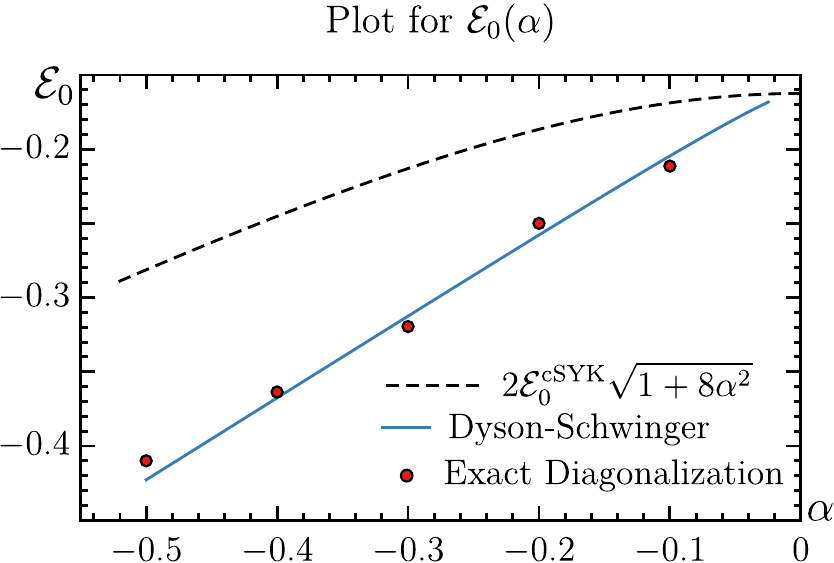}
  \end{center}  
\caption{  
Comparison of the ED and DS calculations of ${\cal E}_0 (\alpha)= \lim_{N\rightarrow \infty} E_0(\alpha)/N$.
They show good agreement even though the ED results are available only up to a moderate values of $N$. 
}
\label{Egroundcomp}
\end{figure}

Another important quantity is the leading term in the ground state energy (\ref{Energygs}), $A(\alpha, N)$, which is expected to grow linearly for large $N$.  In fig. \ref{Eground} we plot
$A(\alpha, N)$ for $\alpha=-0.5$ and $\alpha=-0.2$, and show the fits
\begin{equation}
A (\alpha, N)= {\cal E}_0 (\alpha) N + D(\alpha) + O(1/N)
\ .
\end{equation} 
In fig, \ref{Egroundcomp} we plot
${\cal E}_0 (\alpha)= \lim_{N\rightarrow \infty} E_0(\alpha)/N$ for a range of negative $\alpha$. This shows good agreement with the corresponding calculation using DS equation as a function of $\alpha$. 

\subsection{Line of Fixed Points}
 
Along the fixed line $0\leq \alpha \leq 1$ there is no symmetry breaking, and the large $N$ spectrum is gapless in every charge sector. In fact, for such values of $\alpha,$ near the edge the density of state should behave as 
\begin{equation}
\rho_{2\text{cSYK}}(E)=\int dE' \rho_{\text{cSYK}}(E-E')\rho_{\text{cSYK}}(E')\sim E^2.
\end{equation}

Along the fixed line, we expect the gaps to be of order $1/N$ for excitations of both the $Q_-$ and $Q_+$ charges, so that both $U(1)_-$ and $U(1)_+$ compressibilities are well-defined:
\begin{align}
& E_0= A(\alpha)+ 
\frac{Q_-^2}{2 B_- (\alpha, N)  } + \frac{Q_+^2}{2 B_+ (\alpha, N)  }\ , \notag \\
& B_\pm (\alpha, N)= K_\pm (\alpha) N + C_\pm (\alpha)+ O(1/N)\ .
\end{align}
For $0<\alpha<1$, we find that $B_+> B_-$ for all the values of $N$ we have studied. This leads to the fact that, for odd $N$, the ground state does not have $Q_+=0$. 
Indeed, for odd $N$, the lowest possible values of $(Q_+, Q_-)$ are $(0,\pm 1)$ and $(\pm 1, 0)$. Since $B_+> B_-$, 
there are two ground states with $Q_+=\pm 1, Q_-=0$ for odd $N$. 
On the other hand, for
even $N$ there is a unique ground state with $Q_+= Q_-=0$.  

For $\alpha=0$, $B_-=B_+$. Therefore, the compressibilities are equal: $K_+(0) \approx K_- (0) \approx 2.08$. Indeed, for $\alpha=0$ the Hamiltonian is simply a sum of two cSYK Hamiltonian with the common $J_{ijkl}$, so that for large $N$
\begin{equation}
E_0= A (0)+ 
\frac{1}{2 N K_{\textrm{cSYK}}  } \left (Q_1^2 + Q_2^2 \right ) \ .
\end{equation}
To compare with our normalizations, $K_{\textrm{cSYK}}= K_+ (0)/2 \approx 1.04$. Thus, our finding for $\alpha=0$ is in good agreement with the result $K_{\textrm{cSYK}} \approx 1$ from \cite{Gu:2019jub}.

We have also done fits of the two large $N$ compressibilities along the fixed line. 
Then $K_-$ is found to be smaller than $K_+$. This is in conflict with the DS calculations giving equal values, which may be due to the slow convergence of the ED results to the large $N$ limit. The DS formula $K_+(\alpha)= K_+(0)/\sqrt{1+ 8 \alpha^2}$ predicts the value $\approx 1.96$ at $\alpha=0.125$, and $ \approx 0.94$ at $\alpha=0.7$.
If we a priori assume $K_-=K_+=K(ED)$ in the ED fit for $\alpha>0,$ we obtain results in quite good agreement with the DS calculations. For example, at $\alpha=\frac{1}{8}$, $K(ED)\approx 1.92$ vs $K(DS)\approx 1.96.$ At $\alpha=0.7,$ we get $K(ED)\approx 0.95$ vs $K(DS)\approx 0.94.$

\section{The $U(2)$ symmetric model}
\label{sec:su2}

A special case is $\alpha=1/4$ where the Hamiltonian becomes (\ref{ham2}), and the symmetry is enhanced to $SU(2)\times U(1)_+$.
In this section we assemble various results at this special point, which is interesting because it corresponds to an SYK-like model with a non-abelian global symmetry \cite{Yoon:2017nig, Liu:2019niv, Iliesiu:2019lfc, Kapec:2019ecr}. 

\begin{figure}[h!]
  \begin{center}  
    \includegraphics [width=0.45\textwidth, angle=0.]{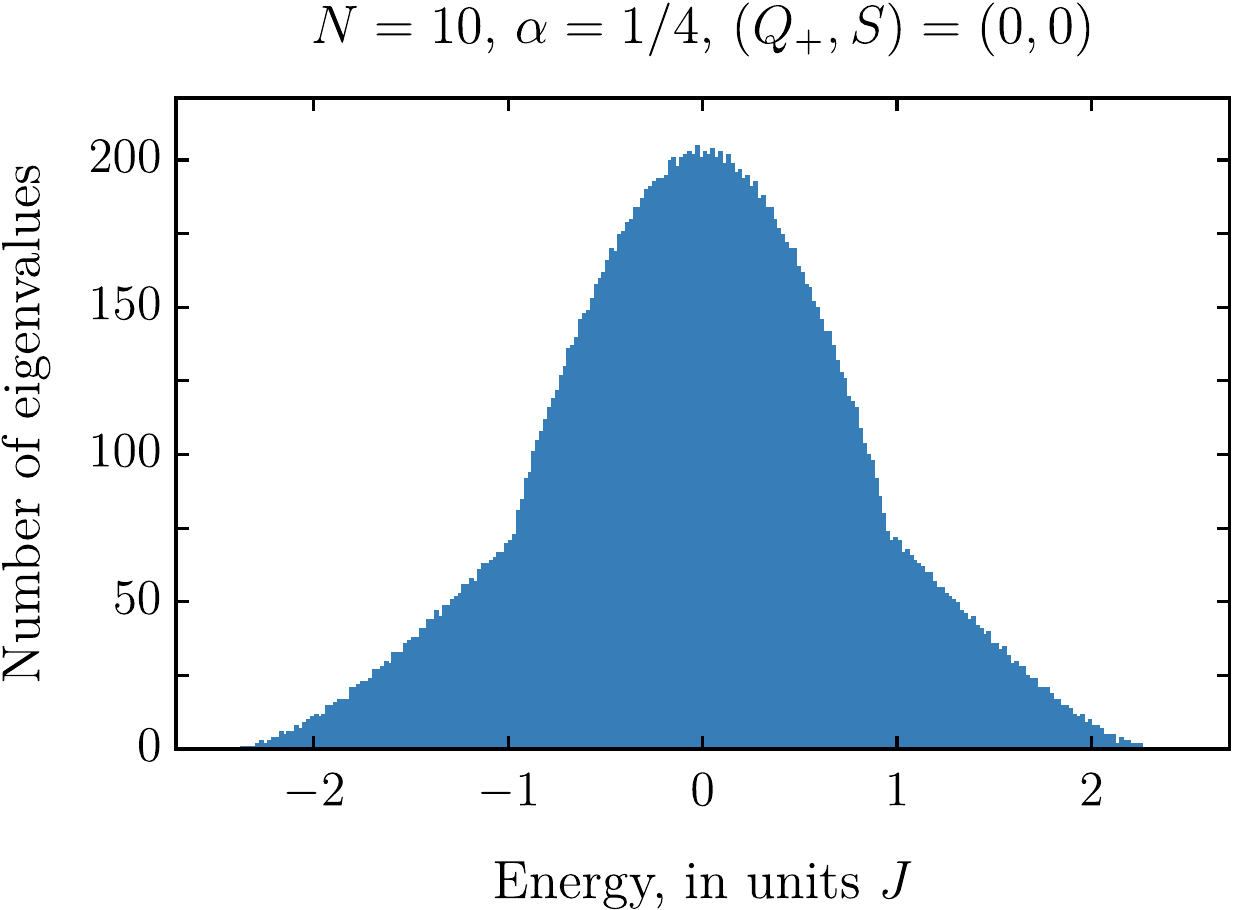}\qquad
        \includegraphics [width=0.45\textwidth, angle=0.]{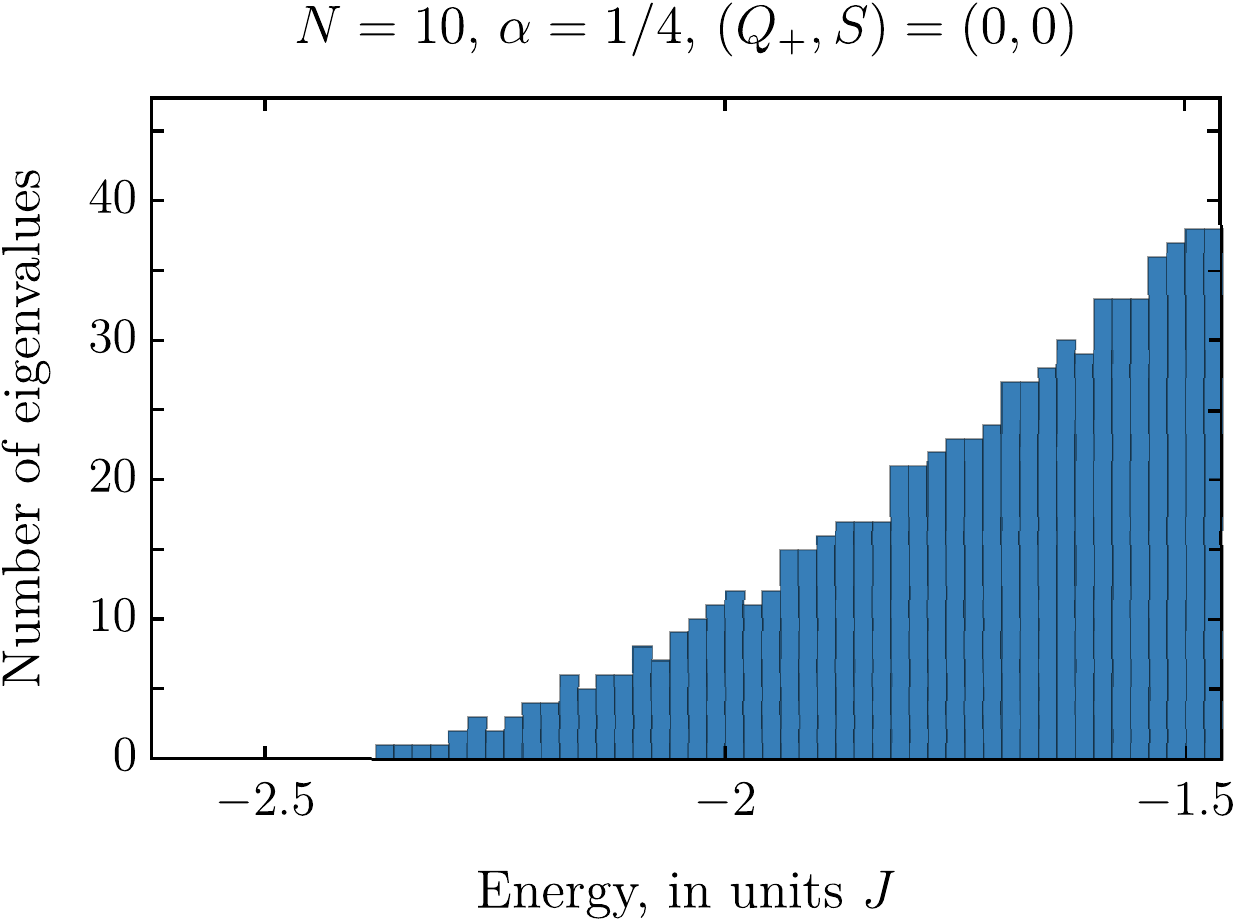}
  \end{center}
  
\caption{ Density of states in the $(Q_{+},S)=(0,0)$ sector for a single realization at the $U(2)$ symmetric point $\alpha=\frac{1}{4}$ for $N=10$. The ground state is in this $U(2)$ invariant sector. On the right we enlarge the region near the ground state; this shows that there is no significant gap.  }
\label{dosU2}
\end{figure}

Due to the $SU(2)$ symmetry, there are some exact degeneracies in the spectrum between states with different values of $Q_-= 2 S_z$. 
The states naturally split into sectors labeled by the $U(1)_+$ charge $Q_+$ and the $SU(2)$ spin $S$. In fig. \ref{dosU2} we show the histogram for the $U(2)$ invariant states,
which have $Q_+=S=0$. Such states appear only when $N$ is even, and the unique absolute ground state is in this sector. The histogram was obtained from a single realization of the Hamiltonian (\ref{pholesymmetric}) with $N=10$, and it shows that the $U(2)$ symmetric theory is in the gapless phase.

Let us discuss the low-energy effective action for the $U(2)$ symmetric theory. We expect that instead, of the $U(1)_+\times U(1)_-$ sigma model, we now have $ SU(2)\times U(1)_+$.  
The low energy effective action for the $SU(2)$ part is:
\begin{align}
& S_{\textrm{SU(2)}} = -\frac{B_{\textrm{SU(2)}}}{4} \int d\tau \Tr \l U^\dagger \pr_\tau U \r^2\ , \notag \\
& B_{\textrm{SU(2)}}= N K_{\textrm{SU(2)}}+ C_{\textrm{SU(2)}} + O(1/N)\ ,
\end{align}
where $U(\tau)$ is a $SU(2)$ matrix variable.
Previously, we obtained compressibilities from
coupling to $\mu_-$ chemical potential. Let us argue that this calculation does not change.
Indeed, corrections  $\propto N$ to the free energy depend on \textit{classical} properties of this sigma-model,
since we have a factor of $N$ in front. Upon introducing a chemical potential for the $U(1)_-$ subgroup of $SU(2)$,
we have to study the following action:
\beq
-\frac{B_{\textrm{SU(2)}}}{4} \int d\tau \Tr \l U^\dagger \pr_t U + \diag(\mu_-,-\mu_-) \r^2 \ .
\eeq
Its contribution to the Gibbs potential is again $- B_{\textrm{SU(2)}} \mu_-^2/2$. We find using the DS equations that 
\beq
K_{\textrm{SU(2)}} \approx 1.7
\label{DScomp}
\ .\eeq

However, the low energy spectrum is very different, as it involves \textit{quantizing} the sigma model.
Namely, now the excitations come in $SU(2)$ multiplets with energies given by a quadratic Casimir of 
$SU(2)$. Namely, for a multiplet with $Q_- /2\in (-S,-S+1,\dots, S)$ the energy is given by:
\begin{equation}
\delta E = \frac{2 S ( S+ 1) }{N K_{\textrm{SU(2)}}+ C_{\textrm{SU(2)}} }\ .
\end{equation}
Therefore, we find for large $N$:
\begin{equation}
E_0\approx A+ \frac{1}{2 (N K_{+}  + C_+)} Q_+^2+ \frac{2}{ N K_{\textrm{SU(2)}}  + C_{\textrm{SU(2)}}} S (S+1)\ ,
\end{equation}
where $S$ is the $SU(2)$ spin.
For even $N$, the unique ground state occurs in the $Q_+= S=0$. For odd $N$, there are two ground states: they are $SU(2)$ singlets and have $Q_+=\pm 1$. 

\begin{figure}[h!]
    \includegraphics [width=0.46\textwidth, angle=0.]{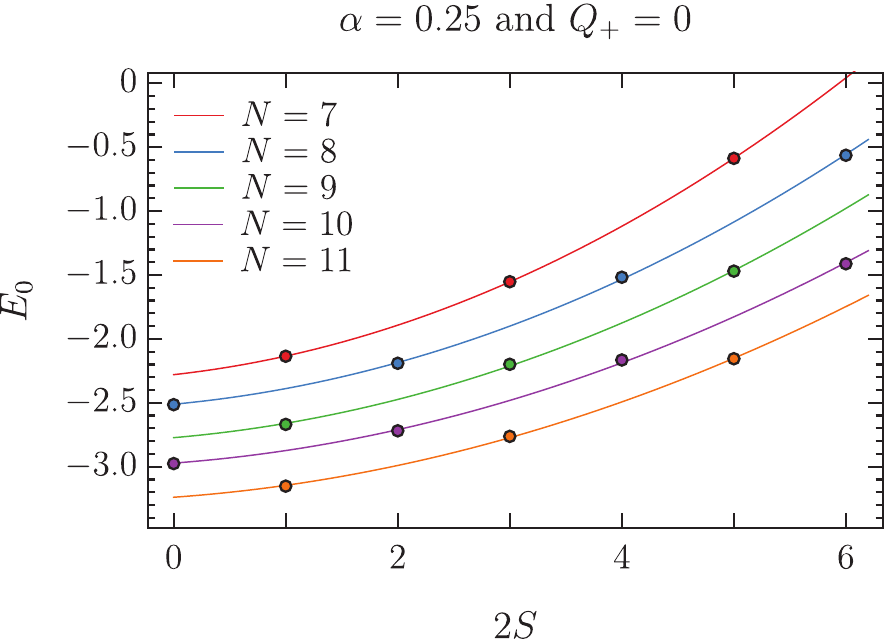} \qquad
    \includegraphics [width=0.46\textwidth, angle=0.]{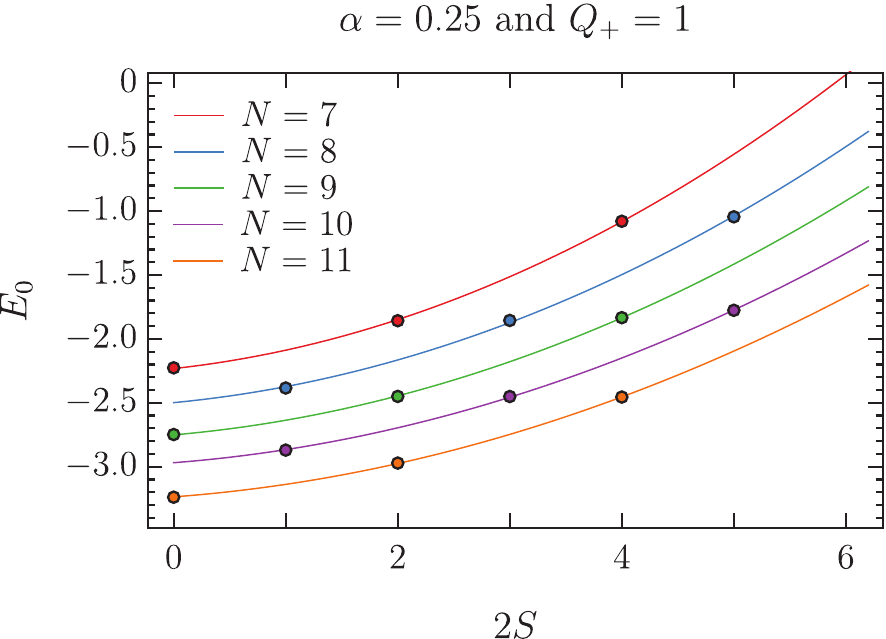} 
       \\
   \\
        \includegraphics [width=0.46\textwidth, angle=0.]{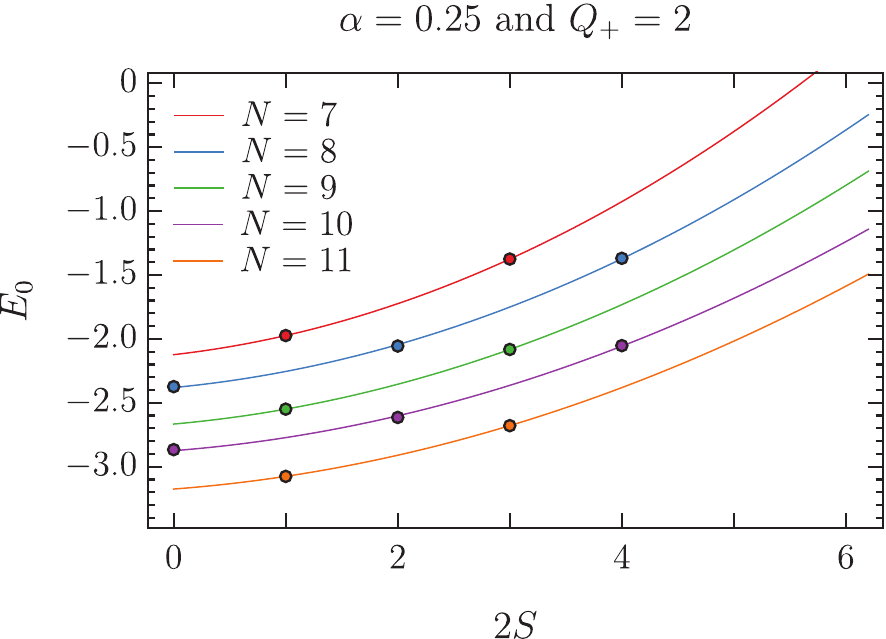}\qquad\;
    \includegraphics [width=0.46\textwidth, angle=0.]{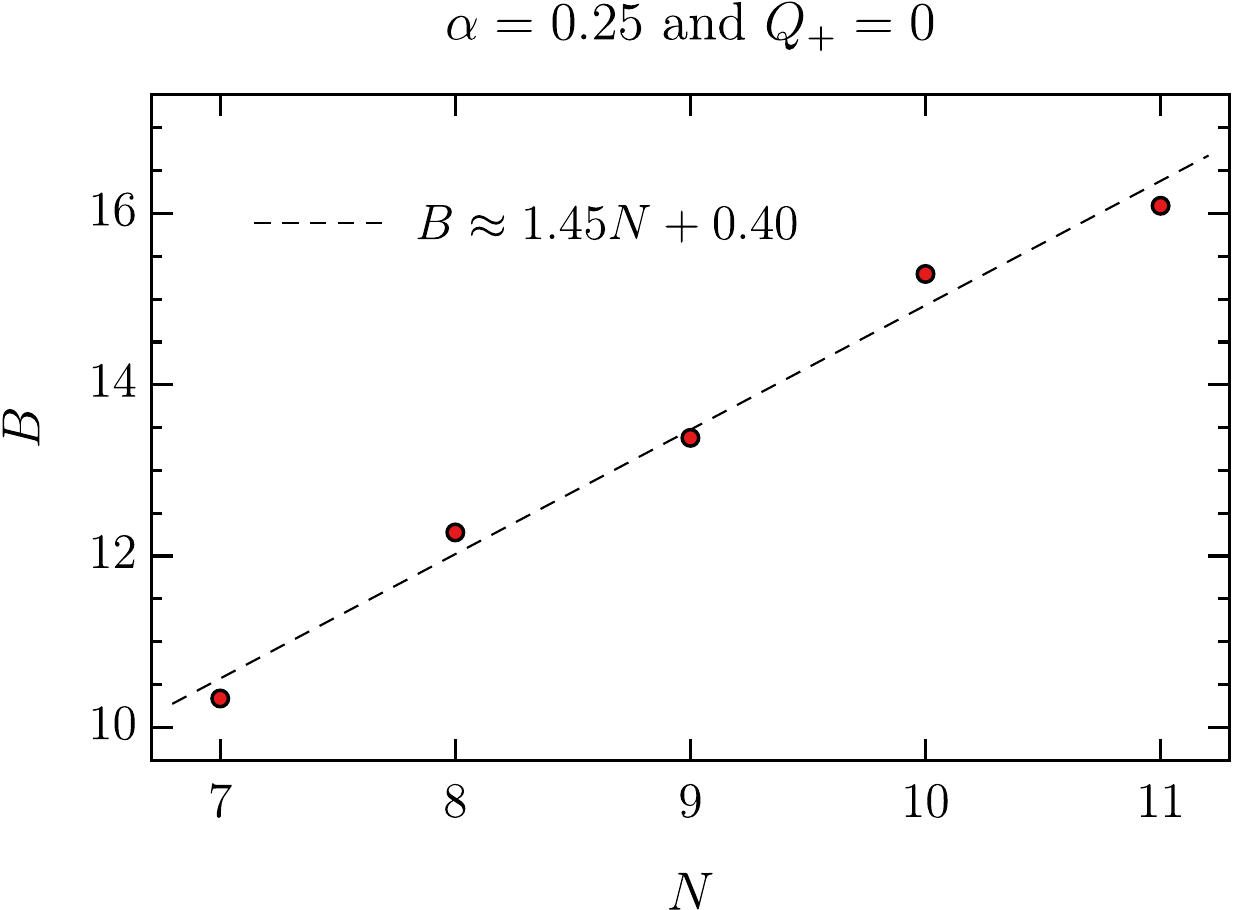}
  
\caption{ The dependence of ground state energy at $\alpha=0.25$ on $SU(2)$ spin $S$ at $Q_+=0,1,2$. 
We use the Ansatz $E_0\approx A+ \frac{1}{2B_+} Q_+^2+ \frac{2}{B_{\textrm{SU(2)}}}  S (S+1)),$ and plot $B_{\textrm{SU(2)}}$ against $N$ to estimate the compressibility, $K_{\textrm{SU(2)}}\approx 1.45.$  }
\label{EQcurvep25}
\end{figure}

A priori, there are two different compressibilities; see fig. \ref{EQcurvep25}. Fitting them separately, we find $K_{\textrm{SU(2)}}\approx 1.45$ and $K_+\approx 1.78$.
The fact that they are different disagrees with the DS results; this could be due to the fact that our data does not access large enough $N$.
However, if we assume that they are equal, then the fit value is $K_+= K_{\textrm{SU(2)}}\approx 1.6$, which is not far from the DS value (\ref{DScomp}).

\section*{Acknowledgments}

IRK dedicates this paper to the memory of his father, Roman Borisovich Klebanov.
IRK and GT are grateful to the Simons Center for Geometry and Physics for hospitality during the workshop "Applications of Random Matrix Theory to Many-Body Physics" in September 2019, where some of this work was carried out. We thank Y. Alhassid, A. Chubukov, S. Giombi, A. Kamenev, J. Kim, J. Maldacena, P. Pallegar and S. Sachdev for useful discussions.
The research of IRK, AM and WZ was supported in part by the US NSF under Grants No.~PHY-1620059 and PHY-1914860.  
The research of GT was supported by DOE Grant No. DE-SC0019030. 

\appendix 

\section{Particle-hole symmetry} 

For the single complex SYK model, the Hamiltonian which respects the particle-hole symmetry $c_i \leftrightarrow c_i^\dagger$, accompanied by 
$J_{ijkl}\rightarrow J^*_{ijkl}$, was given in \cite{Gu:2019jub}
\begin{equation}
H_{cSYK}=\sum_{i,j,k,l=1}^N J_{ijkl} \mathcal{A}\{c_{i}^{\dagger}c_{j}^{\dagger}c_{k}c_{l}\}\ ,
\end{equation} 
where $\mathcal{A}$ denotes total antisymmetrization:
\begin{equation}
\mathcal{A}\{c_{i}^{\dagger}c_{j}^{\dagger}c_{k}c_{l}\}= c_{i}^{\dagger}c_{j}^{\dagger}c_{k}c_{l}
+\frac 1 2 \left (\delta_{i k }c^{\dagger}_{j}c_{l} - \delta_{i l}c^{\dagger}_{j}c_{k} + \delta_{jl }c^{\dagger}_{i}c_{k}-  \delta_{j k}c^{\dagger}_{i}c_{l} +
\frac 1 2 ( \delta_{i l} \delta_{j k} -  \delta_{i k} \delta_{j l} )
\right )\ .
\end{equation}
To make the Hamiltonian of the coupled model, (\ref{ham1}), invariant under the full particle-hole symmetry (\ref{phole}),
we have to add to it similar terms:
\begin{align}\label{EDham}
H_{ed}=&\sum_{i,j,k,l=1}^N J_{ijkl}\bigg(\mathcal{A}\{c_{1i}^{\dagger}c_{1j}^{\dagger}c_{1k}c_{1l}\}+\mathcal{A}\{c_{2i}^{\dagger}c_{2j}^{\dagger}c_{2k}c_{2l}\}
\notag \\
&+8\alpha\left(c_{1i}^{\dagger}c_{2j}^{\dagger}c_{2k}c_{1l}-\frac{1}{2}\delta_{jk}c^{\dagger}_{1i}c_{1l}-\frac{1}{2}\delta^{il}c_{2j}^{\dagger} c_{2k}
+ \frac{1}{4}\delta^{il} \delta_{jk}
\right)\bigg )\ .
\end{align}
 This can also be written as
\begin{align}
H_{ed}=&\sum_{i,j,k,l=1}^N J_{ijkl}\bigg ( c_{1i}^{\dagger}c_{1j}^{\dagger}c_{1k}c_{1l}+c_{2i}^{\dagger}c_{2j}^{\dagger}c_{2k}c_{2l}+8\alpha c_{1i}^{\dagger}c_{2j}^{\dagger}c_{2k}c_{1l}  \notag \\
&+  (1+2\alpha) \left (-2 \delta_{jk}c^{\dagger}_{1i}c_{1l}-2  \delta_{il}c_{2j}^{\dagger} c_{2k}
+ \delta_{il} \delta_{jk} \right )
\bigg ) .
\label{pholesymmetric}
\end{align}
The quadratic and c-number terms 
are subleading in $N$ and thus are not important at large $N.$ They can be important at small $N,$ such as in the exact diagonalizations. 
We note that these terms vanish for $\alpha=-1/2$, so that the original Hamiltonian (\ref{ham1}) is automatically particle-hole symmetric for this value of $\alpha$. 
At another special value, $\alpha=1/4$, the Hamiltonian \ref{pholesymmetric} respects the $U(2)$ symmetry possessed by the purely quartic Hamiltonian (\ref{ham2}).

\section{Zero modes of the quadratic fluctuations}

In this section we give an alternative derivation of the scaling dimension of various primary operators by looking at zero modes of the quadratic fluctuations near the nearly conformal saddle points of the effective action \ref{effaction}.
We assume the time translational invariance and study fluctuations around the symmetric saddle point. The zero modes of the quadratic fluctuation correspond to the operator three point functions $\delta G_{\sigma \sigma'}(\tau)= \langle\frac{1}{N}c^{\dag}_{\sigma i}(\tau) c_{\sigma' i}(0) \mathcal{O}_h(\infty)  \rangle,$ because the DS equations hold up to arbitrary insertion as long as operators are not inserted at $\tau$ or $0. $ In order to not add more contact terms, the operator has to be inserted at $\infty.$ Therefore $\delta G_{\sigma \sigma'}(\tau)$ would correspond to a zero mode in the quadratic fluctuation, and the eigenvector dictates the form of the operator. Note in conformal theory, the 3 point functions between primaries are determined up to a constant 
\begin{equation}
v(\tau)=\langle\frac{1}{N}c^{\dag}_{\sigma i}(\tau) c_{\sigma' i}(0)  \mathcal{O}_h(\infty)  \rangle= \frac{c_{\mathcal{O}}\text{sgn}(\tau)}{|\tau|^{2\Delta-h}}, 
\end{equation}
where $h$ is the scaling dimension of the operator $\mathcal{O}$. In order for the three point function to be non-vanishing, the primary operator $\mathcal{O}$ is necessarily bilinear in the elementary fermions, and a $O(N)$ singlet. Therefore one can use this Ansatz to determine the bilinear operator dimension from the quadratic fluctuation. 
In the following we are going to omit the integrals over $\tau_{1},\tau_{2}$ for brevity.

  We are looking for quadratic fluctuations above the conformal saddle point $G_{*12}=G_{*21}=0$ and $G_{*11}=G_{*22}=G_{*}$,
  where $G_{*}(-\tau)=-G_{*}(\tau)$ and satisfies the Schwinger-Dyson equations
 \begin{align}
 \Sigma_{*}(\tau) = J^{2}(1+8\alpha^{2})G_{*}^{3}(\tau), \quad G_{*}(i\omega_{n})(-i\omega_{n}-\Sigma_{*}(i\omega_{n})) = 1\,.
 \end{align} 
We find for the second variation  
  \begin{align}
\delta^{2}I = \frac{1}{2} G_{*}(\tau_{41})G_{*}(\tau_{23})\textrm{Tr}(\delta \Sigma (\tau_{12})\delta \Sigma(\tau_{34})) -\textrm{Tr}( \delta \Sigma(\tau_{12})\delta G(\tau_{21})) - \frac{J^{2}}{4}\delta^{2}V(G_{ab})\,.
 \end{align} 
 It will be convenient to introduce two vectors 
   \begin{align}
 \delta \textbf{G}(\tau_{12})=(\delta G_{11}, \delta G_{22},\delta G_{12}, \delta G_{21}), \quad \delta \bm{\Sigma}(\tau_{12}) =(\delta \Sigma_{11},\delta \Sigma_{22},\delta \Sigma_{21}, \delta \Sigma_{12})
  \end{align} 
then we find 
 \begin{align}
&\frac{1}{2} G_{*}(\tau_{41})G_{*}(\tau_{23})\textrm{Tr}(\delta \Sigma (\tau_{12})\delta \Sigma (\tau_{34})) = \frac{1}{2}\delta \bm{\Sigma}^{T}(\tau_{12}) G_{*}(\tau_{41})G_{*}(\tau_{23})M \delta \bm{\Sigma}(\tau_{34}), \notag\\ 
&M = \textrm{diag}(\mathbbm{1},\sigma_{x}) \notag\\
&\delta^{2}V = \delta \textbf{G}^{T}(\tau_{12}) G_{*}^{2}(\tau_{12})V\delta \textbf{G}(\tau_{34}), \notag\\
& V = 2 \delta(\tau_{13})\delta(\tau_{24}) \textrm{diag}(
    \mathbbm{1}+ 8\alpha^{2}\sigma_{x} , 8\alpha^{2}\sigma_{x} ) -4 \delta(\tau_{14})\delta(\tau_{23})\textrm{diag}(
    (1+ 4\alpha^{2})\mathbbm{1}+4\alpha^{2}\sigma_{x} , 4\alpha \sigma_{x})
     \end{align} 
where we used that   $G_{*}(-\tau)=-G_{*}(\tau)$. Now we can integrate out fluctuations of $\delta \bm{\Sigma}$ fields 
 and find 
   \begin{align}
\delta^{2}I = -\frac{1}{2}\delta \textbf{G}^{T}(\tau_{12}) \Big(\big(G_{*}(\tau_{32})G_{*}(\tau_{14})\big)^{-1}M+\frac{1}{2}J^{2}G_{*}^{2}(\tau_{12})V\Big)\delta \textbf{G}(\tau_{34})\,.
 \end{align}

 Now let us introduce new variables $\textbf{g}(\tau_{12})=|G_{*}(\tau_{12})| S \delta \textbf{G}(\tau_{12})$, where 
    \begin{align}
S = \frac{1}{\sqrt{2}}\textrm{diag}(
   \sigma_{x}-\sigma_{z} , \sigma_{x}-\sigma_{z}) , \quad S^{T}S = 1\,.
 \end{align}
 In terms of the new variables, the variation corresponds to operators $\{O_2^m, O_1^m, O_4^m, O_3^m\},$ where 
 \begin{equation}
 O_{1,2}^m=c_{1i}^{\dagger}\partial_t^m c_{1i}\pm c_{2i}^{\dagger}\partial_t^m c_{2i} \ ,  \quad O_{3,4}^m= c_{1i}^{\dagger}\partial_t^m c_{2i}\pm c_{2i}^{\dagger}\partial_t^m c_{1i} \ .  \end{equation}
 We can further decompose $\textbf{g}$ into symmetric $\textbf{g}_s(\tau_{12})=\textbf{g}_s(\tau_{21})$ and anti-symmetric $\textbf{g}_a(\tau_{12})=-\textbf{g}_a(\tau_{21})$ sectors under time reflection.
 
 Using the new variables we find 
   \begin{align}
\delta^{2}I =& \frac{3J^{2}(1+8\alpha^{2})}{2} \textbf{g}_{a}^{T}(\tau_{12}) 
 \Big(K_{a}^{-1}\textrm{diag}(1 ,1 , -1 , 1) 
 - \textrm{diag}( \frac{3-8\alpha^{2}}{3 (1+8\alpha^{2})} ,1 ,-\frac{ 8\alpha  (\alpha +1)}{3 \left(1+8\alpha ^2\right)}  ,
 \frac{8 \alpha  (\alpha +1)}{3 (1+8\alpha ^2)})
     \Big)\textbf{g}_{a}(\tau_{34}) \notag\\
  &- \frac{J^{2}(1+8\alpha^{2})}{2} \textbf{g}_{s}^{T}(\tau_{12})  \Big(K_{s}^{-1}\textrm{diag}( 
    1 ,1 , -1 , 1) -  \textrm{diag}( 
    1 ,1 , \frac{8\alpha(\alpha -1) }{1+8\alpha ^2}  , -\frac{8\alpha(\alpha -1)  }{1+8\alpha ^2}) \Big)\textbf{g}_{s}(\tau_{34})\,.
   \end{align}
   where $K_{a}$ and $K_{s}$ are standard SYK kernels
     \begin{align}
&K_{a}(\tau_{1},\tau_{2};\tau_{3},\tau_{4}) = -3 J^{2}(1+8\alpha^{2})|G_{*}(\tau_{12})|G_{*}(\tau_{13})G_{*}(\tau_{24})|G_{*}(\tau_{34})|,  \notag\\
&K_{s}(\tau_{1},\tau_{2};\tau_{3},\tau_{4}) = - J^{2}(1+8\alpha^{2})|G_{*}(\tau_{12})|G_{*}(\tau_{13})G_{*}(\tau_{24})|G_{*}(\tau_{34})|\,.
   \end{align}
The scaling dimensions of the bilinear operators $\{O_1^m, O_2^m, O_3^m, O_4^m \}$ are determined by equating to $1$ the functions (\ref{spec}). 

\section{Diagrammatic Derivation of the Dyson-Schwinger Equations}
\label{DiagSD}

The tensor model Hamiltonian (\ref{tenmod}) has four vertices, which we call 
$v_1,v_2,v_3, v_4.$ We write down Dyson-Schwinger equations for all correlators $G_{\sigma \sigma'}(\tau,\tau')=\frac{1}{N^3}\langle \psi^{\dag,abc}_\sigma (\tau)\psi_{\sigma'}^{abc}(\tau') \rangle,$ allowed by $SU(N)\times O(N) \times SU(N)$ symmetries. Note  $\langle \psi_1^{\dag,abc}(t)\psi^{\dag,abc}_2(0)\rangle$ is forbidden by $SU(N)$. This is the tensor counterpart of using the complex  $J_{ij,kl}$ in the SYK model.

\begin{figure}[h!]
  \begin{center}  
    \includegraphics [width=0.55\textwidth, angle=0.]{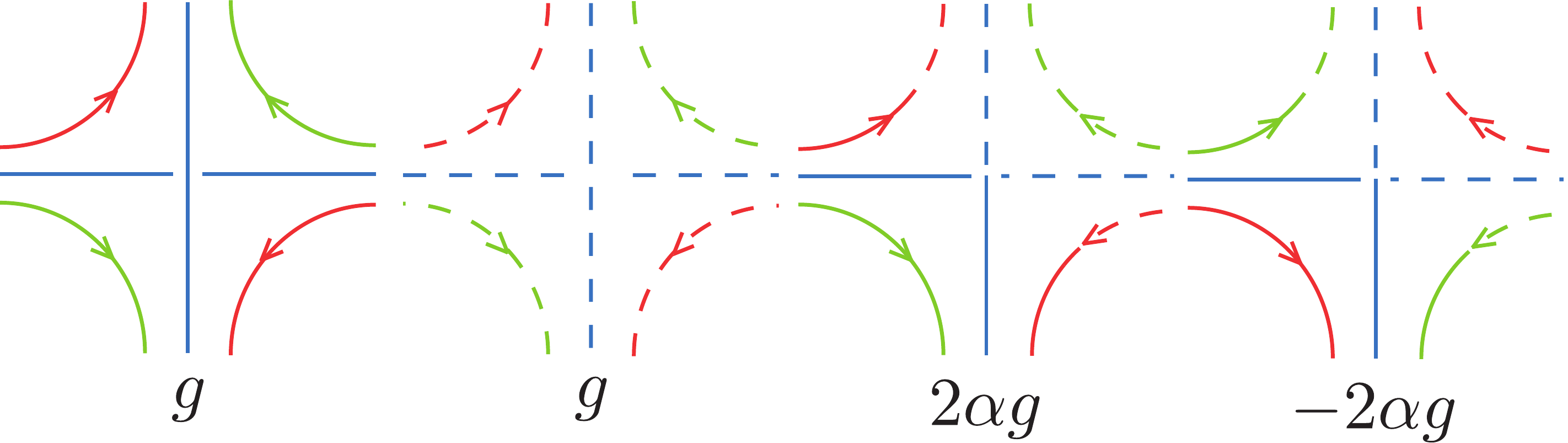}
    \caption{The solid line is for $\psi_1$ and dashed $\psi_2.$}
  \end{center}
\label{vertices}
\end{figure}

At large $N$, only the melonic diagrams contribute to the leading order in $N$. The self-energy can be written in terms of Feynman graphs: 
 \begin{figure}[h!]
  \begin{center}  \label{selfenergy}
    \includegraphics [width=0.75\textwidth, angle=0.]{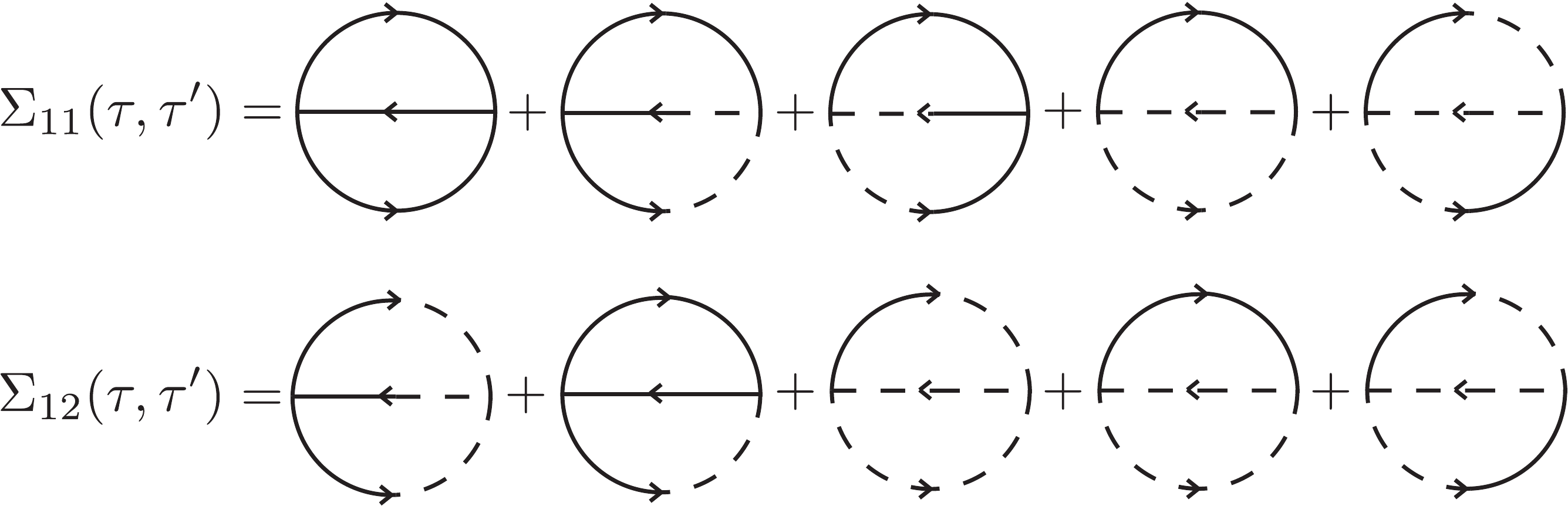}
  \end{center}
\label{vertices}
\end{figure}
Upon drawing the above graphs in the colored line notation, one can check, for example,  
\begin{align}
\Sigma_{11}(\tau)&=-g^2G_{11}(\tau)^2G_{11}(-\tau)-4g^2\alpha G_{11}(\tau)G_{12}(\tau)G_{21}(-\tau)-4g^2\alpha G_{11}(\tau)G_{21}(\tau)G_{12}(-\tau)
\nonumber \\& -2\times (2\alpha g)^2 G_{11}(\tau)G_{22}(\tau)G_{22}(-\tau) -2\times (2\alpha g)^2 G_{12}(\tau)G_{21}(\tau)G_{22}(-\tau)\ ,
\end{align}
which exactly agrees with $\Sigma_{11}$ in Eq.(\ref{secondsdeq}). Similarily $\Sigma_{12}$ agrees. To derive the bilinear spectrum, we shall consider ladder diagrams corrections to 3pt functions along the lines of \cite{Kim:2019upg}, as an effective action at large $N$ is not available for tensor models.

\section{Analytical approximation}
If we assume a particular phase such that the $\mathbb{Z}_4$ symmetry \ref{zfour} is preserved, the  
DS equations can be written using one function $G=\l G_{11}+G_{22} + i \l G_{12} - G_{21}\r \r/2$:
\begin{align}
(-i\omega - \Sigma(\omega)) G(-\omega) = -1\ , \nonumber \\
\Sigma(\tau) = \tilde{J}^2 \l G(\tau)^3 + k G(\tau) G(-\tau)^2 \ ,\r
\label{}
\end{align}
where parameter $k$ is related to $\alpha$ by:
\beq
k= \frac{3+8\alpha+16\alpha^2}{(4\alpha-1)^2}\ ,
\eeq
and $\tilde{J}$ is related to $J$ in the standart formulation with $G_{11}$ and $\alpha$ by
\beq
\tilde{J}^2=\frac{J^2}{4} (1-4\alpha)^2\ .
\eeq
Let us try to use the following ansatz:
\beq
G(\tau) = 
\begin{cases}
a e^{-\mu \tau} + \dots, & \tau>0 \\
-b e^{c \mu \tau} + \dots\ ,& \tau<0\ .
\end{cases}
\label{g:ansatz}
\eeq
We have four unknown constants $\mu,c,a,b$, and the dots indicate faster decaying terms.
The first DS equation can be rewritten in the time domain as:
\beq
\pr_\tau G(\tau) + \int d\tau' \ \Sigma(\tau'-\tau) G(\tau') = \delta(\tau)\ .
\eeq
Evaluating the convolution for $\tau>0$ yields:
\beq
A_1 e^{-\mu \tau} + A_{2+c} e^{-(2+c) \mu \tau} + A_{3c} e^{-3 c \mu \tau}
\eeq
The $A_1, A_{2+c}$ and $A_{3c}$ are easily computed functions of $a,b,c$ and $\mu$:
\beq
A_1 = \frac{a^4 c}{4 (c+1) \mu }+\frac{a^4}{4 (c+1) \mu }-\frac{a^3 b k}{(c+1) \mu }+\frac{a^2 b^2 k}{2 (c+1) \mu }+\frac{a b^3}{\mu -3 c \mu }
\ ,\eeq
\beq
A_{2+c}=\frac{a^3 b k }{(c+1) \mu }+\frac{a^2 b^2 k }{2 (c+1) \mu }\ ,
\eeq
\beq
A_{3c} = \frac{b^4 }{4 c \mu }-\frac{a b^3 }{\mu -3 c \mu }\ .
\eeq

Terms $e^{-2 \mu \tau}, e^{-3c \mu \tau}$ are subdominant and were not present in the ansatz, so we can 
safely ignore them. Therefore we have a single equation:
\beq
\label{eq1}
A_1 = \frac{a \mu}{\tilde{J}}\ .
\eeq 
For $\tau<0$ the convolution equals to:
\beq
B_c e^{c \mu \tau} + B_{1+2c} e^{(1+2c)\mu \tau} + B_{3} e^{3 \mu \tau}\ ,
\eeq
where
\beq
B_c = \frac{a^3 b}{(c-3) \mu }+\frac{a^2 b^2 k}{2 (c+1) \mu }-\frac{a b^3 k}{(c+1) \mu }+\frac{b^4}{4 (c+1) \mu }+\frac{b^4}{4 c (c+1) \mu }\ ,
\eeq
\beq
B_{1+2c} = \frac{a^2 b^2 k }{2 (c+1) \mu }+\frac{a b^3 k}{(c+1) \mu }\ ,
\eeq
\beq
B_3 = \frac{a^4 c}{4 (c+1) \mu }+\frac{a^4}{4 (c+1) \mu }-\frac{a^3 b}{(c-3) \mu }\ .
\eeq

Let us assume that $c>4$. Then we can again ignore the term $e^{(1+2c) \mu \tau}$. However, the term
$e^{3 \mu \tau}$ has to be zero. Therefore we have two equations:
\beq
\label{eq2}
B_3 = 0 \ ,
\eeq
\beq
\label{eq3}
B_c =  \frac{b c \mu}{\tilde{J}}\ .
\eeq
We see that our ansatz is consistent: we managed to eliminate all faster decaying terms.
Moreover, we have 4 unknown variables and only three equations. We will empose one extra condition:
\beq
a+b=1\ .
\label{eq4}
\eeq
If ansatz (\ref{g:ansatz}) were an exact solution, then this condition would have followed from
having a delta function on the right hand side of DS equations.
Unfortunately, (\ref{g:ansatz}) is not an exact solution and at 
very small $\tau$ the faster decaying exponential terms become important. However, we still impose 
eq. (\ref{eq4}) and demonstrate that it agrees with the numerics. So in the end we have four 
algebraic equations (\ref{eq1}), (\ref{eq2}), (\ref{eq3}), (\ref{eq4}) for four unknown variables $a,b,c,\mu$.
This system can be easily solved numerically.

\begin{figure}[h!]
\includegraphics[scale=0.6]{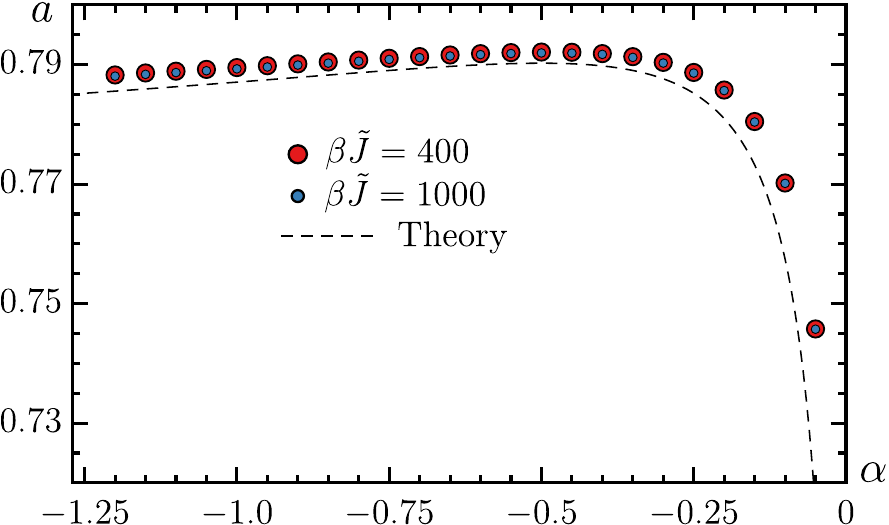}
\includegraphics[scale=0.6]{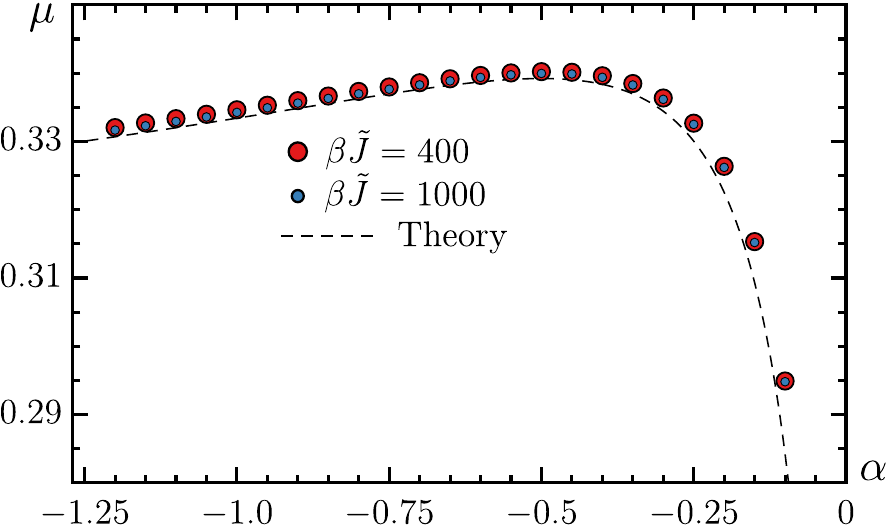}
\includegraphics[scale=0.6]{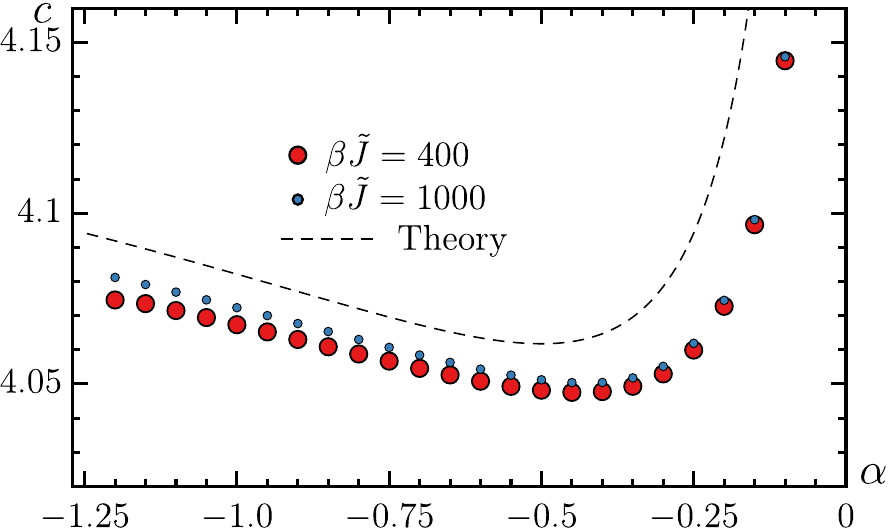}
\caption{$a,\mu,c$ as a function of $\alpha$. Black dashed line: analytical answer obtained by
solving the algebraic system (\ref{eq1}), (\ref{eq2}), (\ref{eq3}), (\ref{eq4}) numerically. 
Dots: numerical solution of DS equations.
Red dots is $\beta \tilde{J} =400$, blue dots is $\beta \tilde{J}=1000$. In both plots $\tilde{J}=1$.}
\label{fig:comparison}
\end{figure}

For comparison, we solve the DS equations numerically for $\beta \tilde{J}=400$(black dots) 
and $\beta \tilde{J}=1000$(red dots) and
fixing $\tilde{J}=1$. After that, we
fit the numerical solution with exponents (\ref{g:ansatz}). This way we obtain numerical
values of $a,b,c,\mu$. The comparison with analytical answer is presented on Figure \ref{fig:comparison}.

Let us note, however, that this approximation does not describe very well the behavior at small Euclidean
times $\tau$. Graphs \ref{Greenfunplot} clearly indicate that $G_{12}$ does not have a linear term near $\tau=0$:
\beq
G_{12} = c_1 - c_2 \tau^2, \ c_1, c_2 > 0\ .
\eeq
Generically, ansatz (\ref{g:ansatz}) does have a linear term near $\tau=0$, by the coefficient 
in front of it is small.

\vspace{10mm}

\bibliographystyle{ssg}
\bibliography{u1citations}

\end{document}